%% file: new_charm_munich_hepex.tex
\documentclass[12pt]{article}
\usepackage{epsfig}

\input econfmacros.tex

\def\Title#1{\begin{center} {\Large {\bf #1} } \end{center}}

\begin{document}

\Title{New charm resonances}

\begin{center}{\large \bf Contribution to the proceedings of HQL06,\\
Munich, October 16th-20th 2006}\end{center}

\bigskip\bigskip


\begin{raggedright}  

{\it Tadeusz Lesiak\index{Lesiak, T.}\\
Institute of Nuclear Physics\\
Polish Academy of Sciences \\
ul. Radzikowskiego 152 \\
31-142 Krak\'{o}w, Poland}
\bigskip\bigskip
\end{raggedright}


\section{Introduction}


In the last five years we have witnessed a renaissance
of charm spectroscopy. Several new charmed states have been 
observed, using data samples collected by so-called 
$B$-factories i.e. $e^+e^-$ storage rings dedicated to the
studies of CP violation in the sector of the beauty quark.
These machines are running essentially at the center-of-mass (CMS)
energy corresponding to the maximum of the $\Upsilon(4S)$
resonance (10.58~GeV/c$^2$).
There are three such accelerators and detectors, which are currently taking data.
The oldest one, which contributed a lot of to the heavy flavour physics
in the past twenty years, is the CLEO apparatus~\cite{Kubota:1991ww,Briere:2001rn}
 at the CESR~\cite{CESR}
 storage-ring (Cornell, USA).
After collecting the data sample of~16~fb$^{-1}$,  
the CLEO collaboration has  moved since 2003 to the lower energy
working point corresponding to the maximum of the $\psi(3770)$.
The other two detectors working at $B$-factories:
 the BaBar~\cite{BABAR} at PEP-II~\cite{PEPII} (Stanford, USA) and Belle~\cite{BELLE} at KEKB~\cite{KEKB}
(Tsukuba, Japan) have collected in the last few
years enormous data samples corresponding to 370~fb$^{-1}$(630~fb$^{-1}$), respectively. The KEKB 
is, in fact,
the record holder as far as the luminosity is concerned with its peak value
 of $1.65\times 10^{34}$~cm$^{-2}$s$^{-1}$.
It is worthwile to stress here that the cross-section for the continuum process 
$e^+e^-\to c\bar{c}$ (1.3~nb) is  
comparable to the one for the reaction $e^+e^-\to\Upsilon(4S)\to B\bar{B}$ (1.1~nb).
As a result, $B$-factories can be safely considered as $c$-factories too.
Moreover, charmed hadrons can he reconstructed here relatively easy, due the `clean'
environment provided by $e^+e^-$ collisions.

This paper is divided into several chapters, each one of them discussing the observation
of an individual new state and entitled with its name. 
The following new meson-like charmed hadrons are talked over:
$X(3872)$, $Y(3940)$, $X(3940)$, $\chi^{\prime}_{c2}(3930)$, $Y(4260)$, $h_c$ 
and the $c\bar{s}$ states $D_{sJ}$.
Then the following new observations of charmed barions will be described: 
$\Sigma_c(2800)$, 
$\Lambda_c(2940)$,
$\Xi_{cx}(2980)$, $\Xi_{cx}(3077)$ and $\Omega_c^{*}$.


\section{X(3872)}

\begin{figure}[hbt]
  \begin{center}
    \includegraphics[height=5.0cm]{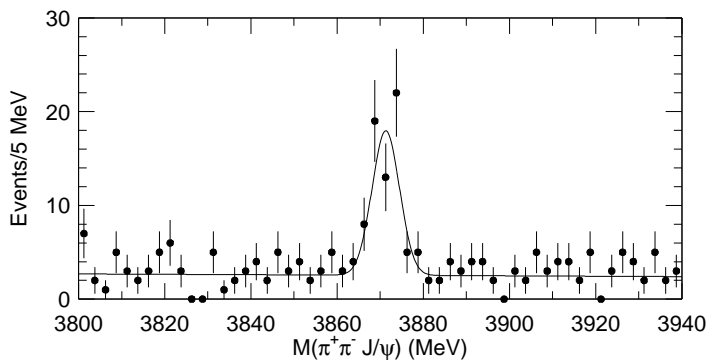}
     \caption{The mass distribution of $J/\psi \pi^+\pi^-$ for the X(3872) resonance, as measured
     by Belle collaboration.}
  \end{center}
\label{MASSX}
\end{figure}

The first new charmed resonance, marked as  X(3872), was discovered by the Belle collaboration
in 2003~\cite{X1BELLE} by analyzing exclusive decays\footnote{charge conjugate
modes are included everywhere, unless otherwise specified.}
$B^+\rightarrow \pi^+\pi^-  J/\psi K^+, J/\psi\to l^+l^-$. 
The $B$ mesons were reconstructed using two kinematical
variables: the energy offset $\Delta E = \sum_i E_i - E_{beam}$
and the beam-constrained
 mass $M_{bc}=\sqrt{E_{beam}^2 - \sum_i (\vec{p_i})^2}$,
where $E_i$ and $\vec{p_i}$  are the center-of-mass (CMS) energies and
momenta of the selected $B$ meson decay products and
$E_{beam}$ is the CMS beam energy.
A very narrow peak in the invariant
mass spectrum of the system $\pi^+\pi^- J/\psi$ was observed (Fig.~\ref{MASSX})
with a statistical 
significance above 10~$\sigma$. The mass of the resonance was determined to be
($3872.0\pm 0.6\pm 0.5$) MeV/c$^2$ and a width
below 2.3 MeV (90\% C.L.), which is consistent with the detector resolution.

The observation of $X(3872)$ was very quickly confirmed by the CDF~\cite{XCDF},
D0~\cite{XD0} and BaBar~\cite{XBABAR} experiments.
At first glance the $X(3872)$ would appear as an ideal candidate for one of the, unobserved yet, 
charmonium states. Among $(c\bar{c})$ states, the ones expected to be closest in mass to X are those
belonging to the multiplets $1D$ and $2P$ multiplets~\cite{CC1}--\cite{CC4}.
However, it soon turned out that the, discussed below, properties of none of these
states are in agreement with measured properties of $X(3872)$.
This fact stimulated the development of several theoretical models assuming the exotic nature
of this new resonance. 
In particular, the coincidence of the X mass with the 
$D^0\bar{D^{*0}}$ threshold i.e. ($3871.3\pm 1.0$)~MeV/c$^2$ has prompted many
theoretical speculations that X(3872) may be a so-called deuson~\cite{XDEUSON1}--\cite{XDEUSON4} 
 i.e. a loosely bound molecular state of these two mesons
or a tetraquark i.e. a tightly bound open charm diquark-antidiquark
state~\cite{XTETRAQ,FAUSTOV1}.
Other models attributed the $X(3872)$ as a $(c\bar{c})$-gluon hybrid meson~\cite{XHYBRID}, a glueball
with a $(c\bar{c})$ admixture~\cite{XGLUEB} or the so called threshold cusp effect~\cite{XCUSP}.

\begin{figure}[tb]
\begin{minipage}[b]{.5\linewidth}
\setlength{\unitlength}{1mm}
  \begin{picture}(70,40)
  \put(12,30){\bf a)}
    \includegraphics[height=3.9cm]{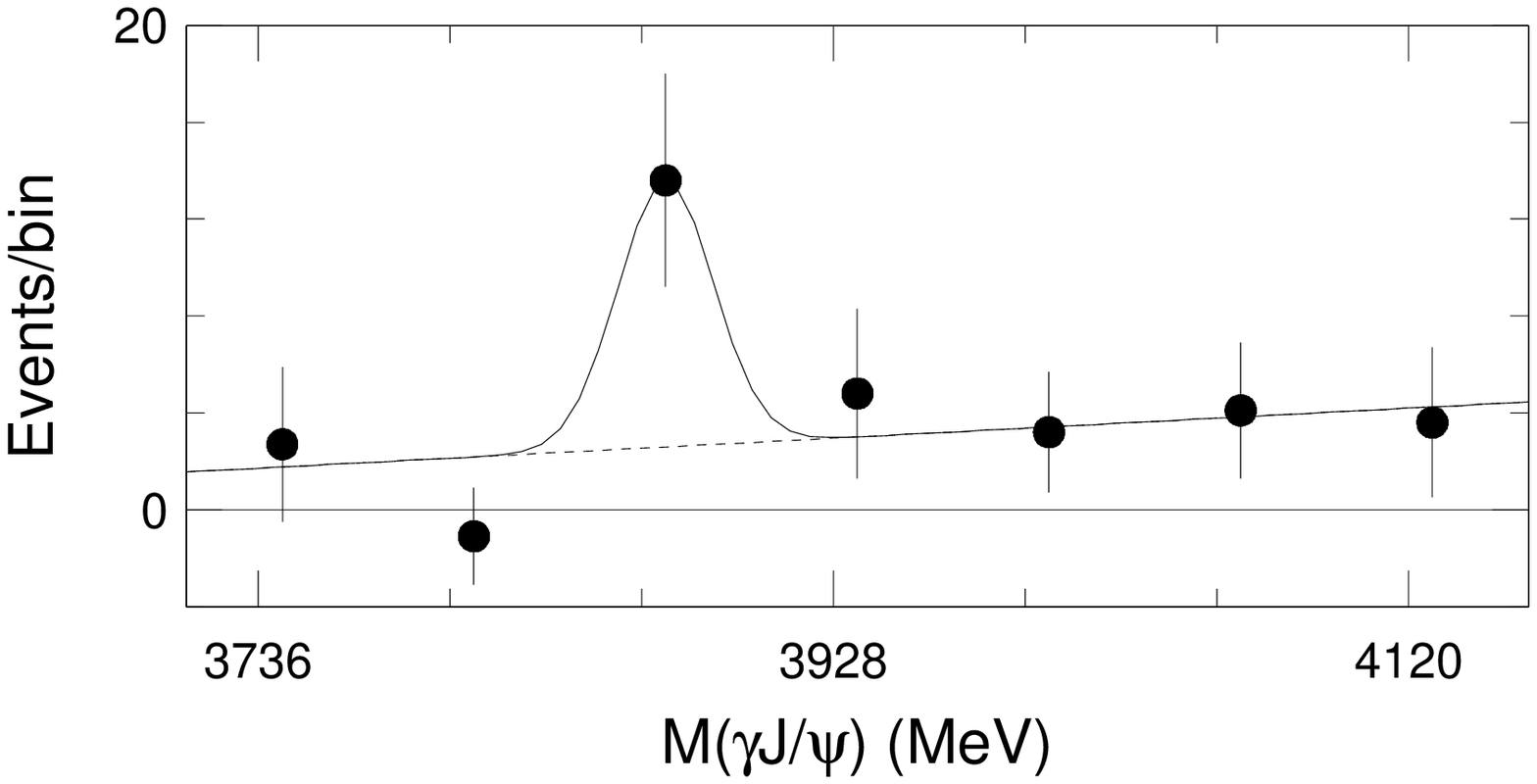}
  \end{picture}
\end{minipage}\hfill
\begin{minipage}[b]{.5\linewidth}
\setlength{\unitlength}{1mm}
  \begin{picture}(70,40)
  \put(12,30){\bf b)}
    \includegraphics[height=3.9cm]{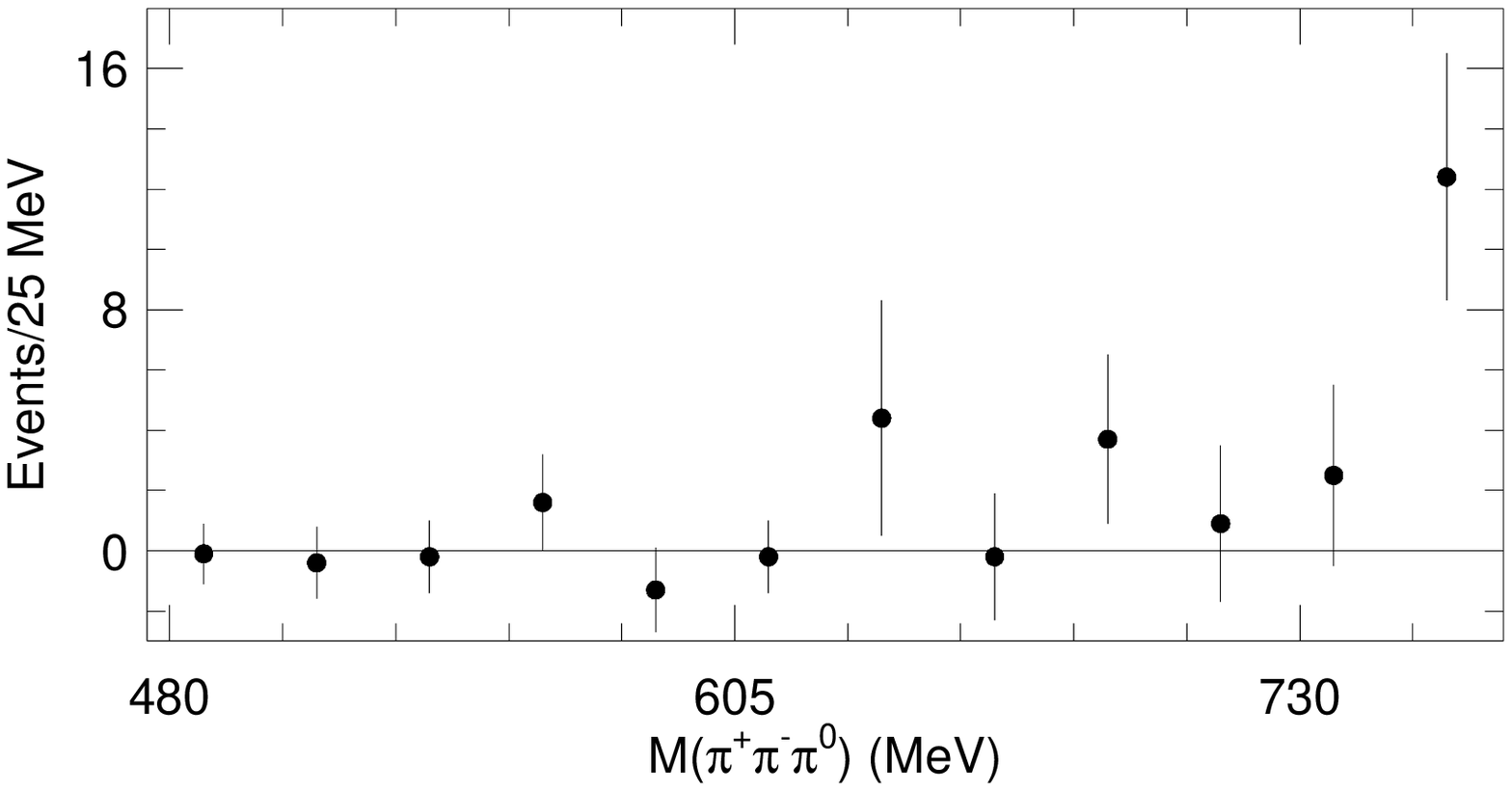}
  \end{picture}
\end{minipage}
  \caption{The yield of $B$ mesons from the decay {\bf a)}
             $B^0\rightarrow \gamma J/\psi K$,
            in bins of the  $\gamma J/\psi$ invariant mass and {\bf b)}
            $B^0\rightarrow \pi^+\pi^-\pi^0 J/\psi K$,
            in bins of the  $\pi^+\pi^-\pi^0$ invariant mass,
  determined by the Belle collaboration
from fits to the $\Delta E$ and $M_{bc}$ distributions.} 
\label{GAMM3PI}
\end{figure}

The Belle collaboration,
 has also provided the first evidence for two
new decay modes of the $X(3872)$: $X\rightarrow \gamma J/\psi$
and $X\rightarrow \pi^+\pi^-\pi^0 J/\psi$~\cite{X0505037},
 observed in exclusive $B$ meson decays to the final states
$\gamma J/\psi K$ and $\pi^+\pi^-\pi^0 J/\psi K$, respectively.
The yield of the decay $B\rightarrow \gamma J/\psi K$ plotted in bins
of the $\gamma J/\psi$ invariant mass (Fig.~\ref{GAMM3PI}{\bf a)})
 exhibits an excess of 
$13.6\pm 4.4$ events (statistical significance of 4$\sigma$).
This evidence was was recently confirmed by the BaBar collaboration~\cite{X0607050}
with the signal yield of $19.2\pm 5.7$ events (3.4$\sigma$).
The observation of this decay establishes unambiguously that the 
charge-conjugation parity of the X(3872) is positive and indicates the presence of the
$c\bar{c}$ component in its wave function.
The partial width ratio 
$\Gamma(X\rightarrow \gamma J/\psi)/\Gamma(X\rightarrow \pi^+\pi^- J/\psi)$
amounts to $0.14\pm 0.05$. This result is, in particular, in contradiction with the 
$\chi_{c1}^{\prime}$ ($1^{++}$ charmonium) assignment for X as in this 
case a value around 40 would be expected. The second decay mode 
$X\rightarrow \pi^+\pi^-\pi^0 J/\psi$ was found to be dominated by the
sub-threshold decay $X\rightarrow \omega^* J/\psi$. This is motivated by the
fact that the yield
of $B$ mesons plotted in bins of the $\pi^+\pi^-\pi^0$ invariant mass
(Fig~\ref{GAMM3PI}{\bf b)}) inside of the signal region from the decay
$X\rightarrow\pi^+\pi^-\pi^0 J/\psi$ is consistent with zero except
for the $M(\pi^+\pi^-\pi^0)>750$~MeV/c$^2$.
There, the excess of $12.4\pm 4.1$ events (4.3$\sigma$) is observed.
The ratio of branching fractions
{\cal B}($X\to\pi^+\pi^-\pi^0 J/\psi$)/{\cal B}($X\to\pi^+\pi^- J/\psi$).
was measured to be  $1.0\pm 0.4\pm 0.3$, which implies a large
violation of isospin symmetry. This in turn points at the presence of both 
$u\bar{u}$ and $d\bar{d}$ pairs in the X wave function.
The overall properties of the above two decays are in reasonable agreement
with the $D^0\bar{D}^{0*}$ molecule hypothesis.

The Belle collaboration also attempted  to determine the $J^{PC}$
quantum numbers of the X(3872)~\cite{X0505038} by studying 
the angular distributions
of the decay $X\to\pi^+\pi^- J/\psi$,
as suggested by J.L.~Rosner~\cite{ROSNER}\index{Rosner, J.L.}. 
Among the twelve possible $J^{PC}$ assignments, half
 ($0^{--}$, $0^{+-}$, $1^{--}$, $1^{+-}$, $2^{--}$ and $2^{+-}$) 
 may be discarded due to their negative charge conjugation-parity.
The assignments  $0^{-+}$  and $0^{++}$ are strongly disfavoured by the
analysis of angular distributions.
The additional two odd-parity possibilities: 
$1^{-+}$ and $2^{-+}$  are discarded as for them the 
dipion invariant mass spectrum is expected
to be much softer to compare with the data.
The above considerations leave only two assignments: $1^{++}$ and $2^{++}$
as the possible $J^{PC}$ of $X$.
The decay angular distributions and  $\pi^+\pi^-$ angular distribution agree
well with the $1^++$ hypothesis.

The assignment $2^{++}$ was  disfavoured by the recent observation 
by Belle~\cite{X0606055} of a near-threshold enhancement in the
 $D^0\bar{D^0}\pi^0$ invariant mass  in $B\to KD^0\bar{D^0}\pi^0$ decays.
It corresponds to $23.4\pm 5.6$ 
 signal events (6.4$\sigma$) at mass ($3875.4\pm 0.7 \pm 1.1$) MeV/c$^2$ which
is around two standard deviations higher than the world average for $X(3872)$~\cite{PDG}. 
Taking for granted that the observed near-threshold enhancement is 
due to the $X(3872)$,  the decay of a spin 2
state to  three pseudoscalars ($D^0\overline{D^0}\pi^0$) 
would require at least one pair of them to be in a relative D wave.
In such a configuration the near threshold production would be
strongly suppressed by a centrifugal barrier.

\begin{figure}[tb]
 \begin{center}
\setlength{\unitlength}{1mm}
    \begin{picture}(100,70)(0,-7)
     \includegraphics[height=6.0cm]{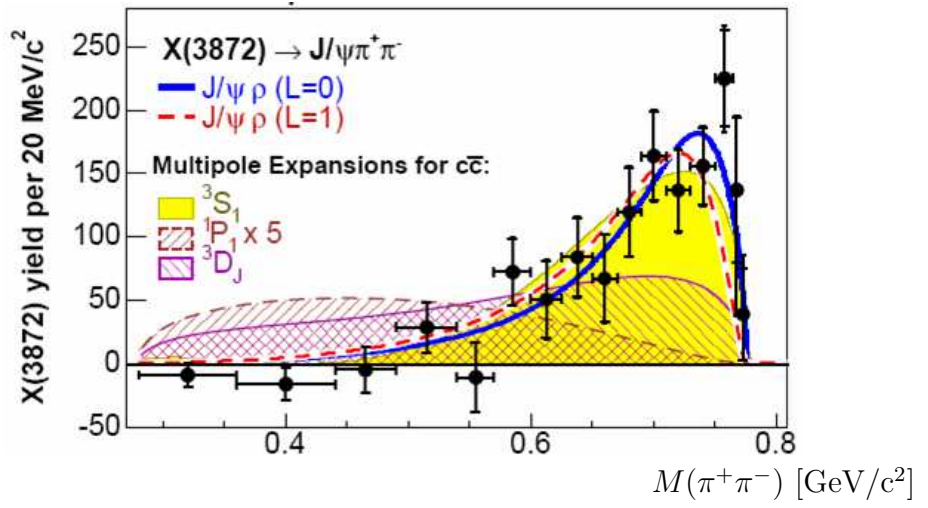}
     \put(-20,-5){$M(\pi^+\pi^-)$ [GeV/c$^2$]}
    \end{picture}
  \caption{The dipion mass spectrum for the $X(3872)$ (data points), as measured by the
CDF collaboration,  together with fits to different $J^{PC}$ hypotheses.}
 \end{center}
\label{MCDFPIPI}
\end{figure}

The CDF collaboration~\cite{XCDFPIPI} has recently studied 
the spin-parity of $X(3872)$ using a high-statistics sample
of $\approx 3000$ events of $X(3872)\to\pi^+\pi^- J/\psi$.
The shape of the 
$\pi^+\pi^-$ invariant mass distribution was compared
with the predictions corresponding to all relevant $J^{PC}$ values.
(Fig.~\ref{MCDFPIPI}). It was found that both $1^{++}$ and $2^{-+}$
assignments fit reasobably the data.

Collecting the above information it seems the most plausible that X(3872)
is a deuson. This conjecture is supported in particular by the pattern
of its decay modes and the favoured spin-parity assignment $1^{++}$.


\begin{figure}[tbh]
 \begin{center}
  \includegraphics[height=5.0cm]{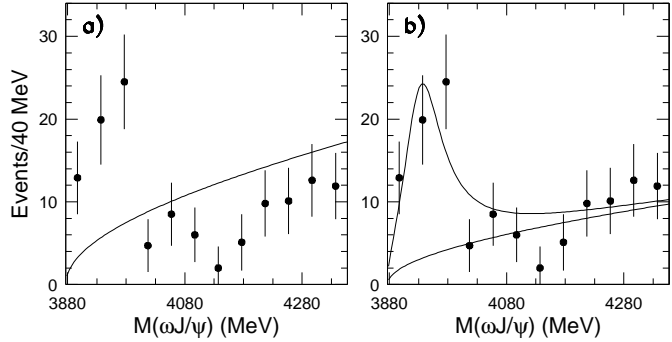}
 \end{center}
  \caption{   $B^+\to K^+ \omega J/\psi$ signal yields vs  $M(\omega J/\psi)$ as 
determined by the Belle collaboration. The curve in (a) shows the result of a 
fit that includes only a phase-space-like threshold function. The curve in (b)
 corresponds to the result of a fit that includes an $S$-wave Breit-Wigner resonance term.} 
\label{YYYY}
\end{figure}


\section{Y(3940)}


In 2004 The Belle collaboration  observed another
new state, denoted as Y(3940) and produced in    $B^+\to K\omega J/\psi$ decays~\cite{Y3940}.
In this study events with  $M(K\omega)<1.6$ GeV/c$^2$ were rejected in order to
remove the contribution from $K^*\to Kw$ decays.
A fit to the $\omega J/\psi$ invariant-mass distribution (Fig.\ref{YYYY}) yielded
a signal of $58\pm 11$ events (8.1$\sigma$) corresponding to a mass of
 ($3943\pm 11\pm 13$) MeV/c$^2$ and the width
($87\pm 22\pm 26$) MeV. 

Both mass and width of $Y$  are in agreement with expectations for a radially 
excited charmonium state $\chi_{cJ}^{\prime}$. This interpretation is also 
strengthened by the observation of the corresponding
($b\bar{b}$) decay $\chi_{b1}^{\prime}\to \omega\Upsilon(1S)$~\cite{CHIB}.
Such a ($c\bar{c}$) state would, however, decay 
predominantly to $D\bar{D}^{(*)}$ pairs, which is not observed.
Moreover, for the $\chi_{cJ}^{\prime}$ hypothesis one would expect that
${\cal B}(B\to K\chi_{cJ}^{\prime}) < {\cal B}(B\to K\chi_{cJ})=4\times 10^{-4}$.
Taking into account the value of the product 
 ${\cal B}(B\to K Y) {\cal B}(Y\to\omega J/\psi) = (7.1\pm 1.3\pm 3.1)\times 10^{-5}$ ,
determined by Belle, this implies that ${\cal B}(Y\to \omega J/\psi) > 12$~\%.
Such a value would seem exceptionally high for any charmonium state with a mass
above $D\bar{D}^{(*)}$ threshold.

The above drawbacks of the conventional charmonium interpretation of $Y$, in particular
the lack of its decay to $D\bar{D}^{(*)}$ and a large ${\cal B}(Y\to\omega J/\psi)$, are in fact
advantages while taking for granted the hypothesis of a $c\bar{c}$-gluon hybrid~\cite{YHYB1}.
This is also supported by the lattice QCD calculations~\cite{YHYB2} which indicate that a partial
width for the decay to $K \omega J/\psi $ are comparable to the value measured by Belle.
However, the masses of $c\bar{c}$-gluon  mesons predicted by these 
calculations~\cite{YHYB2}--\cite{YHYB4} 
are between 4300 and 4500 MeV/c$^2$ i.e. substantially higher than the measured value.


\section{X(3940)}

\begin{figure}[t]
 \begin{center}
  \includegraphics[height=5.0cm]{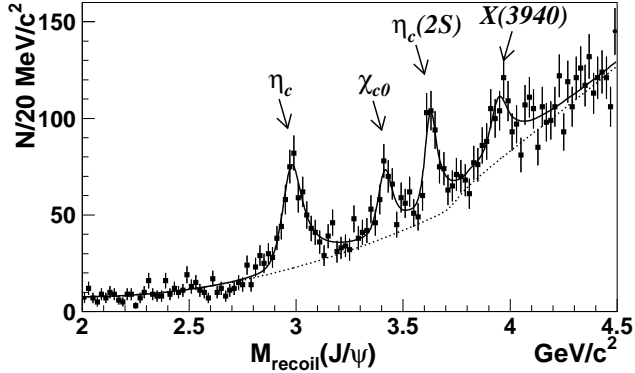}
 \end{center}
  \caption{The distribution of masses recoiling against the reconstructed $J/\psi$
measured by the Belle collaboration
in inclusive $e^+e^- \to J/\psi X$ events. The four enhancements, from left to right,
 correspond to the $\eta_c$, $\chi_{c0}$, $\eta_c(2S)$ and a new state $X(3940)$.}
\label{Y3940}
\end{figure}

Yet another particle, marked as $X(3940)$ with 
the mass of 3940 MeV/c$^2$ was observed by the Belle collaboration in the
process $e^+e^-\to J/\psi X$~\cite{BY3940}. Its signal was seen in the
spectrum of the $J/\psi$ recoil mass (Fig.~\ref{Y3940}) defined as
$M_{recoil}(J/\psi)=\sqrt{(E_{CMS}-E^*_{J/\psi})^2-(cp^*_{J/\psi})^2}/c^2$, where  
$E_{CMS}$ is the  center-of-mass energy of the event and $E^*_{J/\psi}$ ($p^*_{J/\psi}$)
denote the CMS energy (momentum) of the $J/\psi$, respectively.
The previous studies of this process reveiled the presence of three states:
$\eta_c$, $\chi_{c0}$ and $\eta_c(2S)$. The new analysis, using significantly
higher statistics, provided the observation of the fourth 
particle  in the $J/\psi$ recoil mass spectrum.
Its mass  was estimated to be $(3943\pm 6\pm 6)$~MeV/c$^2$ and 
and the width smaller than 52 MeV (90~\% C.L.). 
The search for $X(3940)$ decay modes yielded the evidence for 
$X\to D^*\bar{D}$ (${\cal B}=96^{+45}_{-32}\pm 22$~\%).
No signal was observed for $X\to D\bar{D}$ (${\cal B}<41$~\% (90~\% C.L.))
and $X\to \omega J/\psi$ (${\cal B}<26$~\% (90~\% C.L.))
The properties of X(3940) match the expectations~\cite{CC4} 
of the $3^1S_0$ charmonium state, denoted also as $\eta_c^{''}$.

It is appropriate to stress here that, in spite the same mass measured,
it is extremely unlikely that the states  $X(3940)$ and $Y(3940)$ coincide.
The X(3940) decays to  $D\bar{D^*}$ and does not decay
to  $\omega J/\psi$. for the Y(3940) the situation is reversed, as far as the
above-mentioned decays are concerned.

\begin{figure}[tb]
\begin{minipage}[b]{.5\linewidth}
\setlength{\unitlength}{1mm}
  \begin{picture}(70,50)
    \includegraphics[height=5.0cm,width=6.5cm,trim=1 20 1 1]{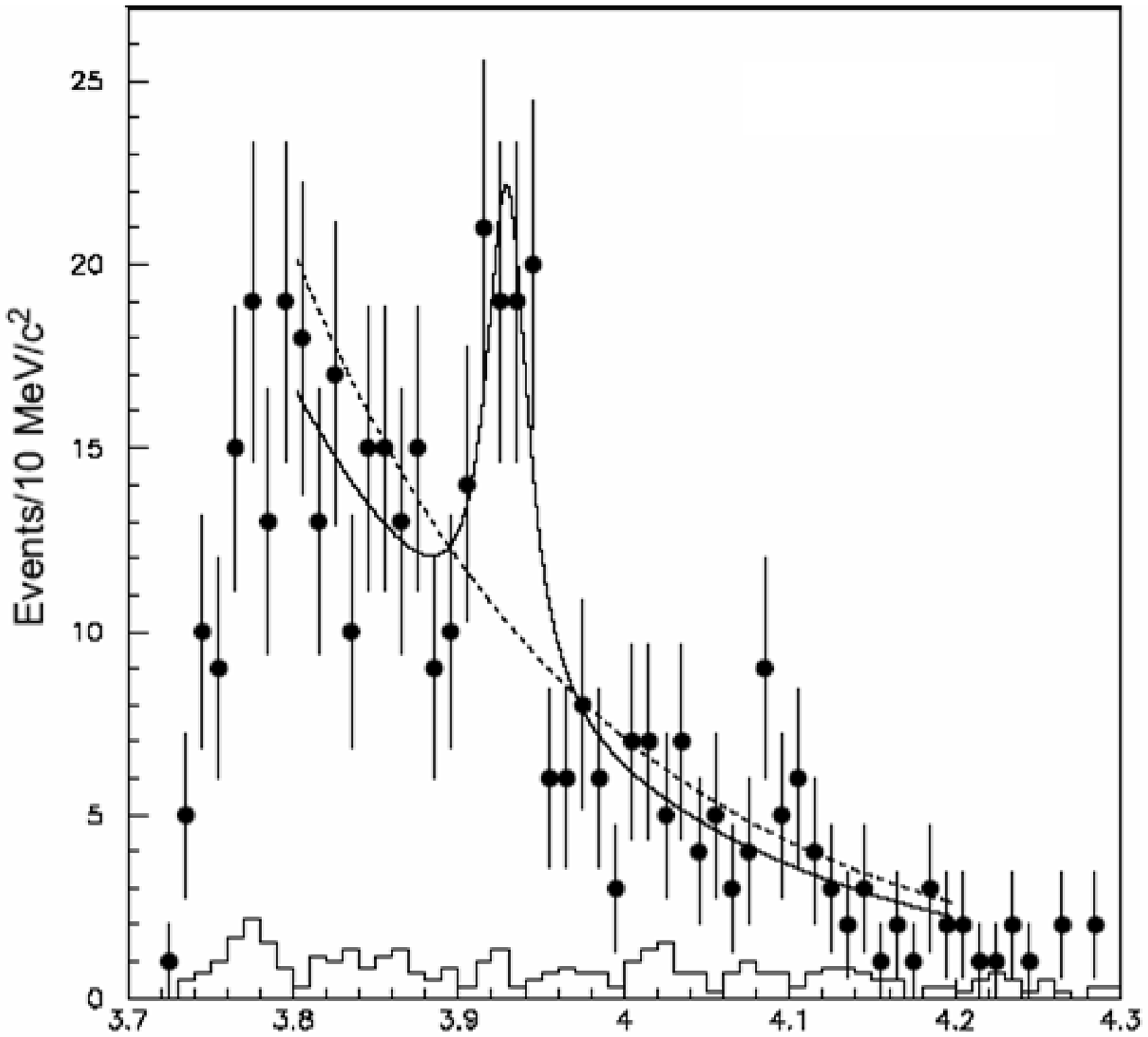}
  \put(-15,44){\bf (a)}
  \put(-40,-3){\bf $M(D\bar{}D)$ [GeV/c$^2$]}
  \end{picture}
\end{minipage}\hfill
\begin{minipage}[b]{.5\linewidth}
\setlength{\unitlength}{1mm}
  \begin{picture}(70,50)
    \includegraphics[height=5.1cm,width=5.5cm]{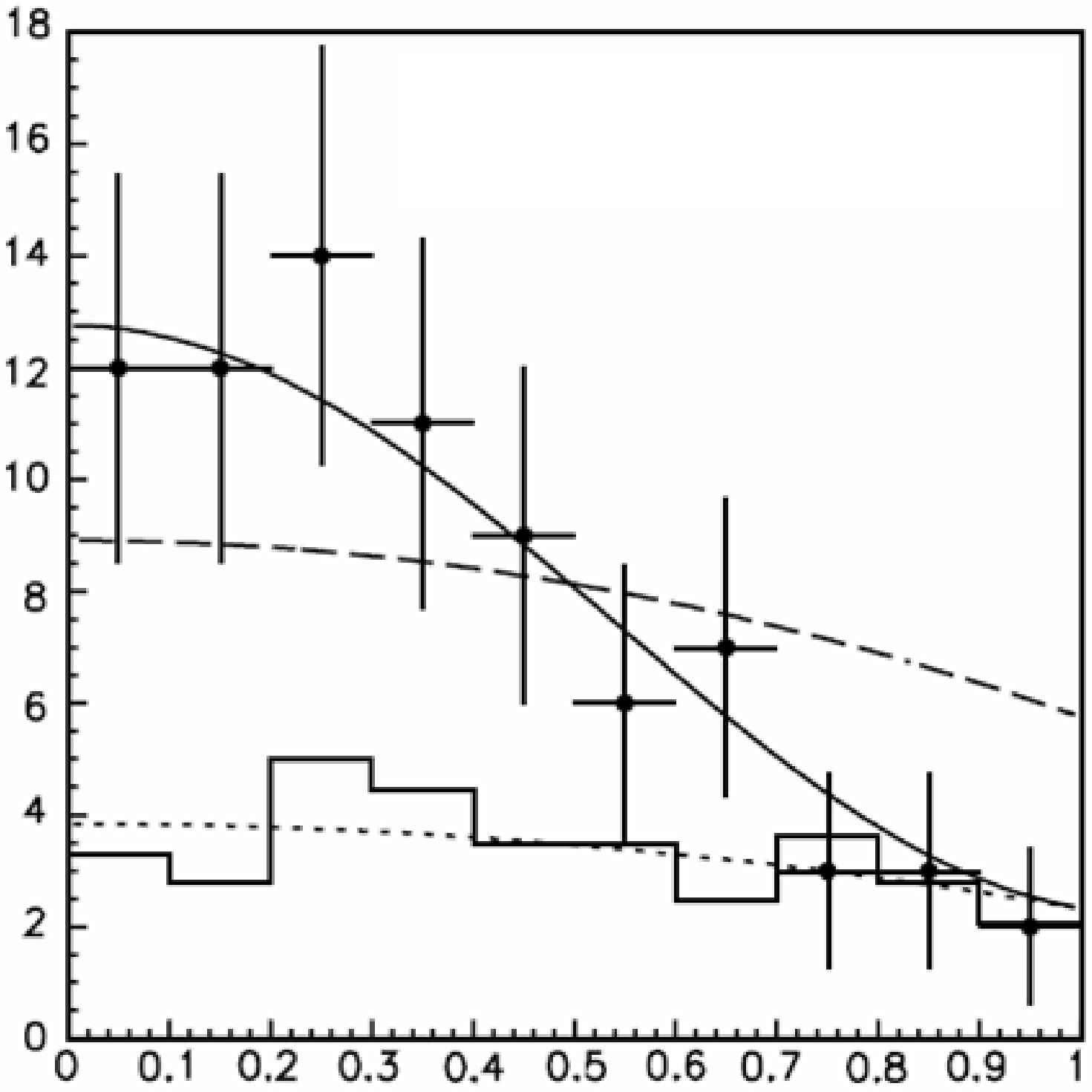}
  \put(-15,44){\bf (b)}
  \put(-30,-3){\bf $| \cos\theta^*|$}
  \end{picture}
\end{minipage}
  \caption{{\bf (a)} Invariant mass distribution of $D\overline{D}$ pairs produced
in two photon processes as measured by Belle. The solid (dashed) curve shows
the fits with (without) a resonance component. The histogram
corresponds to the distribution of the events from the $D$-mass sidebands.
{\bf (b)} The distribution of the angle $\theta^*$ of a $D$ meson relative to the beam axis in the
$\gamma\gamma$ CMS frame. The data points correspond to the  $3.91 < M(D\bar{D})< 3.95$~MeV/c$^2$ region.
The solid histogram shows the yield of background scaled from $M(D\bar{D})$ sidebands.
The solid and dashed curves represent expectations for spin-2 and spin-0 hypotheses, respectively.
The dotted curve interpolates the non-peak background.} 
\label{ZZZZ}
\end{figure}


\section{$\chi^{\prime}_{c2}(3930)$}


The Belle collaboration has also performed the search
 for the production of new resonances in the process $\gamma\gamma\to D\overline{D}$~\cite{ZZZZ}.
Here the two-photon processes are studied in the ``zero-tag'' mode, where the final state
electron and positron are not detected and the transverse momentum of the $D\bar{D}$ system is very small.
The analysis yielded an  observation of a new state, marked as $Z(3930)$,  at the mass and width of ($3929\pm 5\pm 2$) MeV/c$^2$
and  ($29\pm 10\pm 2$) MeV, respectively (Fig.~\ref{ZZZZ} a)). The 
statistical significance of the signal amounted to  5.3$\sigma$. The product of the two-photon radiative width
and branching fraction for the decay to $D\bar{D}$  was found to be
$\Gamma_{\gamma\gamma}\times {\cal B}(Z(3930)\to D\bar{D}) = (0.18\pm 0.05\pm 0.03)$~keV.
The properties of this new state match the
expectations~\cite{CC4,ZSWANSON} 
for the radially excited ($c\bar{c}$) states
$\chi_{c0}^{\prime}$  and
$\chi_{c2}^{\prime}$. 
A study of angular distribution of the $D$ mesons  in the $\gamma\gamma$ CMS frame
showed that that spin-2 assignment is strongly favoured over the spin-0 hypothesis.
Thus the state $Z(3930)$ can be safely interpreted as the $\chi_{c2}^{\prime}$ $2^3P_2$ charmonium.


\begin{figure}[p]
 \begin{center}
  \includegraphics[height=5.0cm]{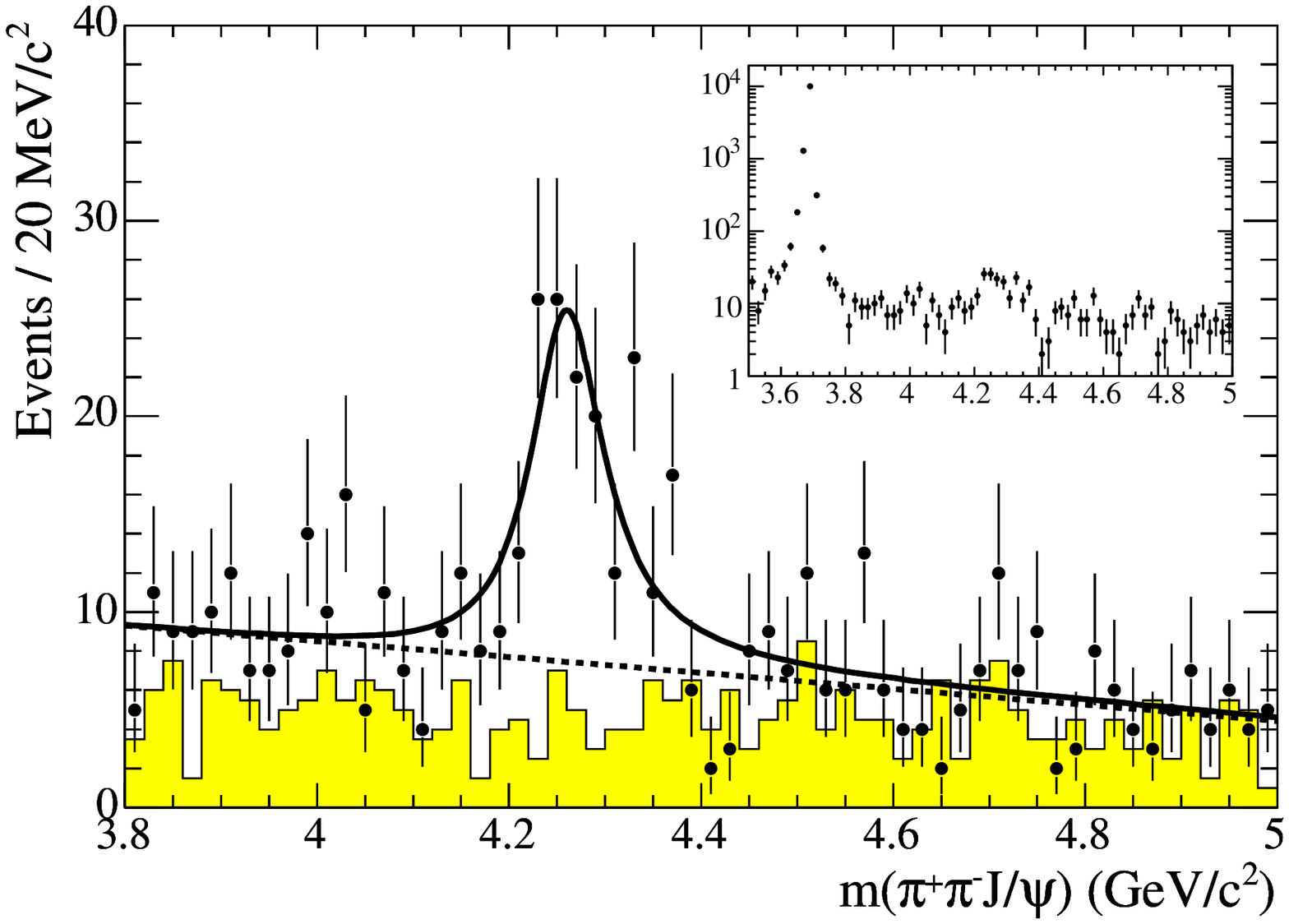}
 \end{center}
  \caption{The $\pi^+\pi^- J/\psi$ invariant mass spectrum measured by BaBar in the range 3.8--5.0 GeV/c$^2$
and (inset) over a wider range that includes the $\psi(2S)$. The points represent the data and
the shaded histogram corresponds to the scaled data from the $J/\psi$ mass sidebands.
The solid line shows the result of the single-resonance fit. The dashed curve represents
the background component.}
\label{FIG1_Y4260}
 \begin{center}
  \includegraphics[height=9.0cm]{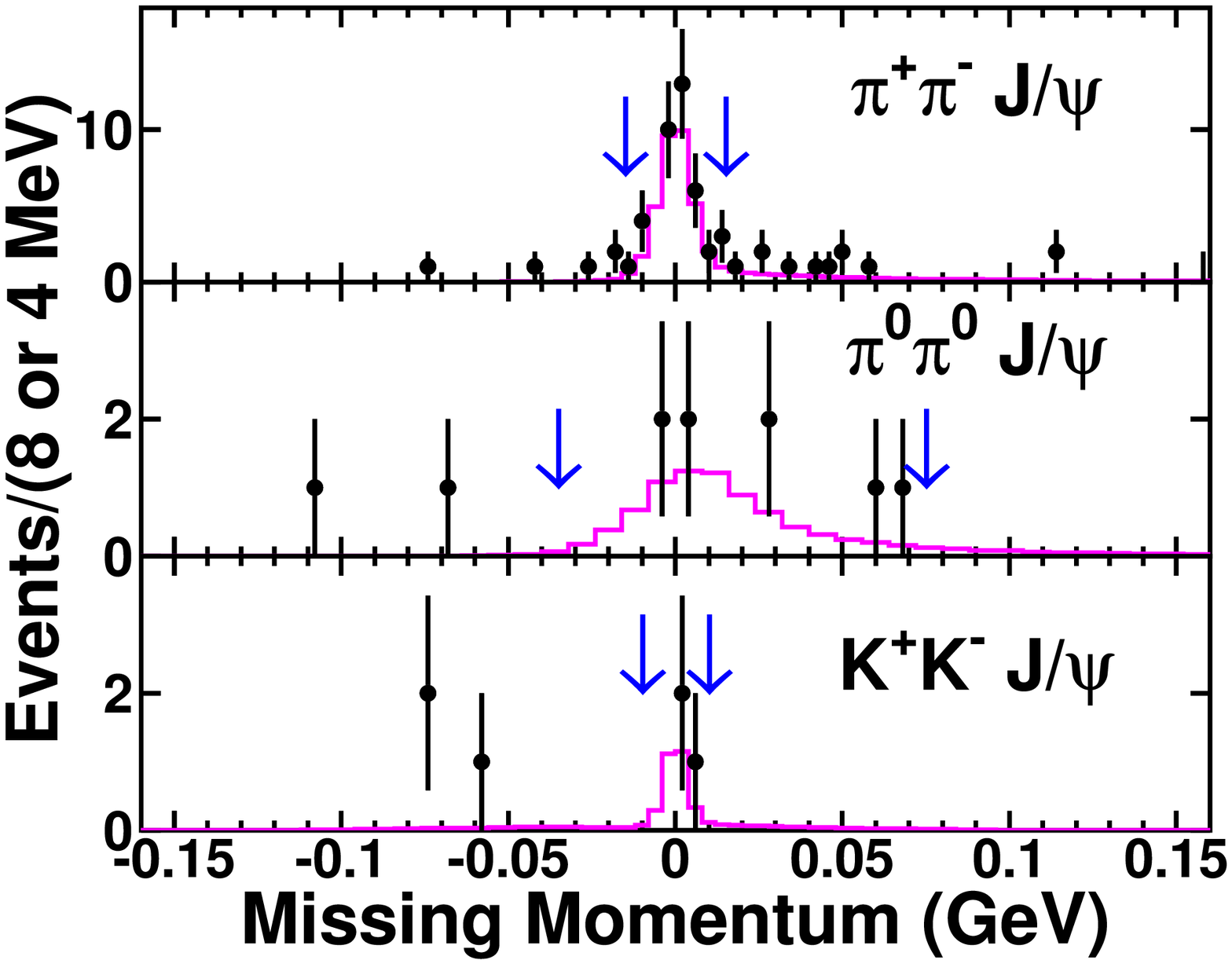}
 \end{center}
  \caption{The missing momentum distribution measured by Cleo collaboration for
 $\pi^+\pi^- J/\psi$ (top),  $\pi^0\pi^0 J/\psi$ (middle) and  
 $ K^+ K^- J/\psi$ (bottom) in the data at $\sqrt{s}=4.26$~GeV (data points)  and
the signal shape as predicted by MC simulation (histogram) scaled to the  net signal size.} 
\label{FIG2_Y4260}
\end{figure}



\section{$Y(4260)$}


\begin{table}[b]
\begin{center}
\begin{tabular}{l|ccc}  
             &  BaBar &  CLEO &  Belle (preliminary)  \\ \hline
 Yield (significance) &   $125\pm 23$ ($>8 \sigma$)  &   $14.1^{+5.2}_{-4.2}$ ($4.9\sigma$)  & $165^{+24+7}_{-22-23}$ ($>7\sigma$)  \\
 Mass (MeV/c$^2$)     &  $4259\pm 8 ^{+2}_{-6}$      &  $4283^{+17}_{-16}\pm 4$              &  $4295\pm 10 ^{+11}_{-5}$ \\ 
 Width (MeV)          &  $88\pm 23 ^{+6}_{-4}$       & $70^{+40}_{-25}\pm 5$                 & $133^{+26+13}_{-22-6}$    \\  
\hline
\end{tabular}
\caption{The parameters of the $Y(4260)$ resonance, as measured by BaBar, CLEO and Belle.}
\label{TAB_Y4260}
\end{center}
\end{table}

The BaBar collaboration has studied initial-radiation (ISR) processes~\cite{Y4260} 
$e^+e^-\to\gamma_{ISR}\pi^+\pi^- J/\psi$ and observed a broad resonance 
in the invariant mass spectrum of $\pi^+\pi^- J/\psi$ near 4.26 GeV/c$^2$(Fig.~\ref{FIG1_Y4260}).
The photon radiated from an initial $e^+e^-$ collision is not detected directly.
Since the new state, marked as $Y(4260)$, is produced in ISR events,
its spin-parity is well defined as $1^{--}$.
The existence of $Y(4260)$ was soon confirmed by CLEO~\cite{Y4260_CLEO}
and Belle~\cite{Y4260_BELLE} collaborations.
The relevant parameters of this new state are collected in Table~\ref{TAB_Y4260}.
It is worthwile to note that the values measured so far by three collaborations are only 
marginally consistent.

The CLEO collaboration has also provided the first observation of $Y(4260)\to\pi^0\pi^0 J/\psi)$
($5.1\sigma$) and found the first evidence for $Y(4260)\to K^+K^- J/\psi)$ ($3.7\sigma$)~\cite{Y4260_CLEO}
(Fig.~\ref{FIG2_Y4260}). Simultaneously, the $e^+e^-$ cross-sections at $\sqrt{s}=4.26$ GeV
were determined for $\pi^+\pi^- J/\psi$ and $\pi^0\pi^0 J/\psi$ final states to be
($58^{+12}_{-10}\pm 4$)~pb and
($23^{+12}_{-8}\pm 1$)~pb, respectively.

The observation of the $\pi^0\pi^0 J/\psi$ contradicts the hypothesis
that $Y$ is a $\chi_{cJ}\rho$ molecule~\cite{Y4260_MOLEC}. The
interpretation of Y as a baryonium state~\cite{Y4260_BARYO} 
is strongly disfavoured by the fact that the $\pi^0\pi^0 J/\psi$ rate is 
about half that of $\pi^+\pi^- J/\psi$. The $Y(4260)$ is located at the dip 
in $R(e^+e^-\to hadrons)$. Similar drop of the cross-section was also found by Belle in the
  exclusive reaction $e^+e^-\to D^{*+}D^{*-}$, measured
 as a function of the CMS energy using  ISR events~\cite{Y4260_PAKHL}.
This dip could be accomodated as a result of $\psi(3S)-\psi(4S)$ interference,
provided that $Y(4260)$ can be interpreted as the conventional
charmonium state $\psi(4S)$~\cite{Y4260_CCBAR}. 
Then, however, the $\psi(3S)$ should exhibit
a substantial coupling to $\pi^+\pi^- J/\psi$, which is not observed.
Two other viable models describe the $Y(4260)$ as a tetraquark~\cite{Y4260_TETRA}
or a $c\bar{c}$-gluon hybrid meson~\cite{Y4260_CCG1}--\cite{Y4260_CCG3}.
The unambiguos interpretation of $Y(4260)$ can be possibly obtained
as a result of careful studies of its open-charm decays, in particular
those with $D$ meson (both $S$ and $P$ wave) pairs.

\begin{figure}[p]
 \begin{center}
  \includegraphics[height=7.8cm]{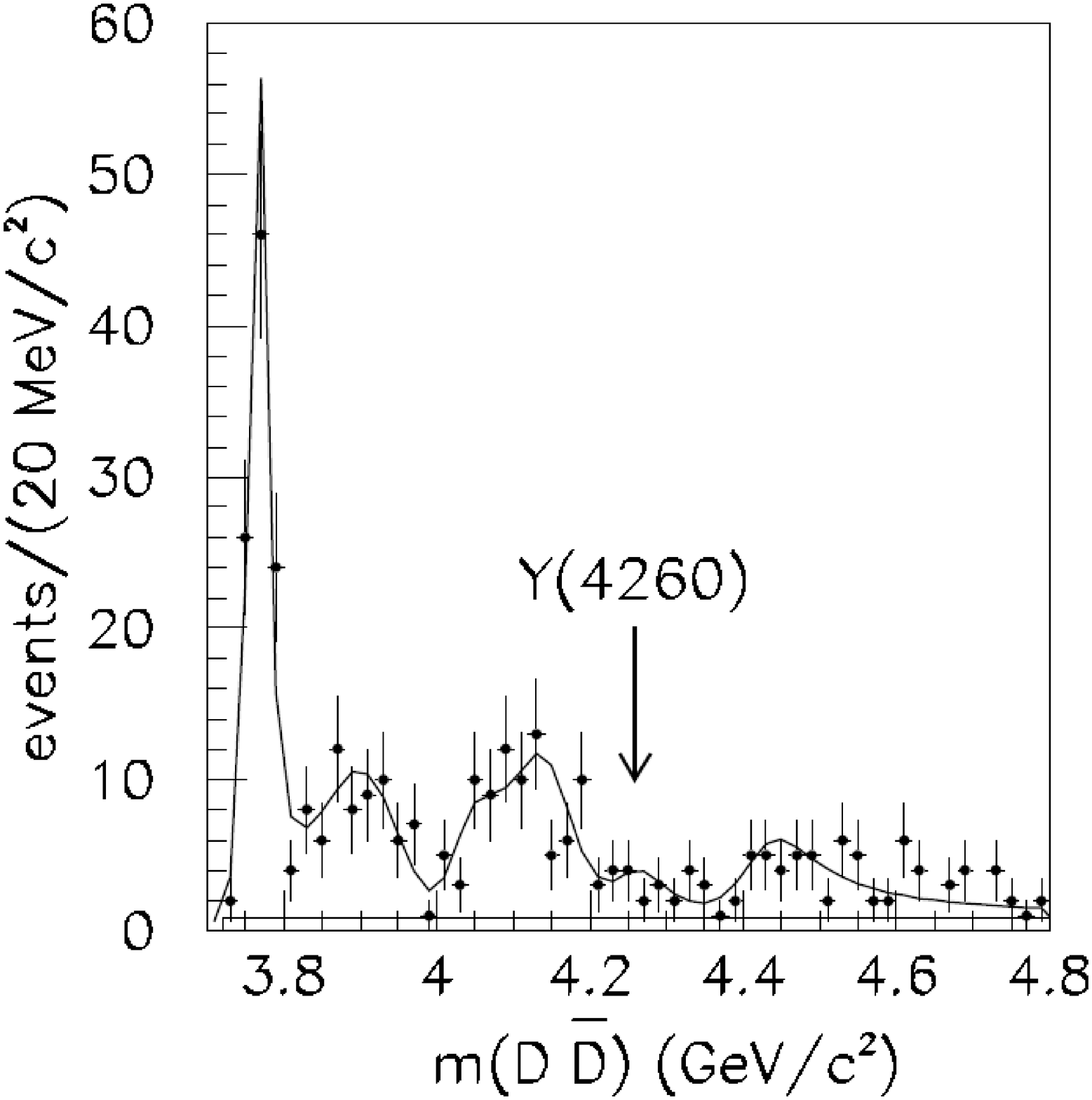}
 \end{center}
  \caption{The $D\bar{D}$ mass spectrum for the ISR sample, as measured
by BaBar. The arrow indicates the expected position of the $Y(4260)$.}
\label{YDD}
 \begin{center}
  \includegraphics[height=7.8cm]{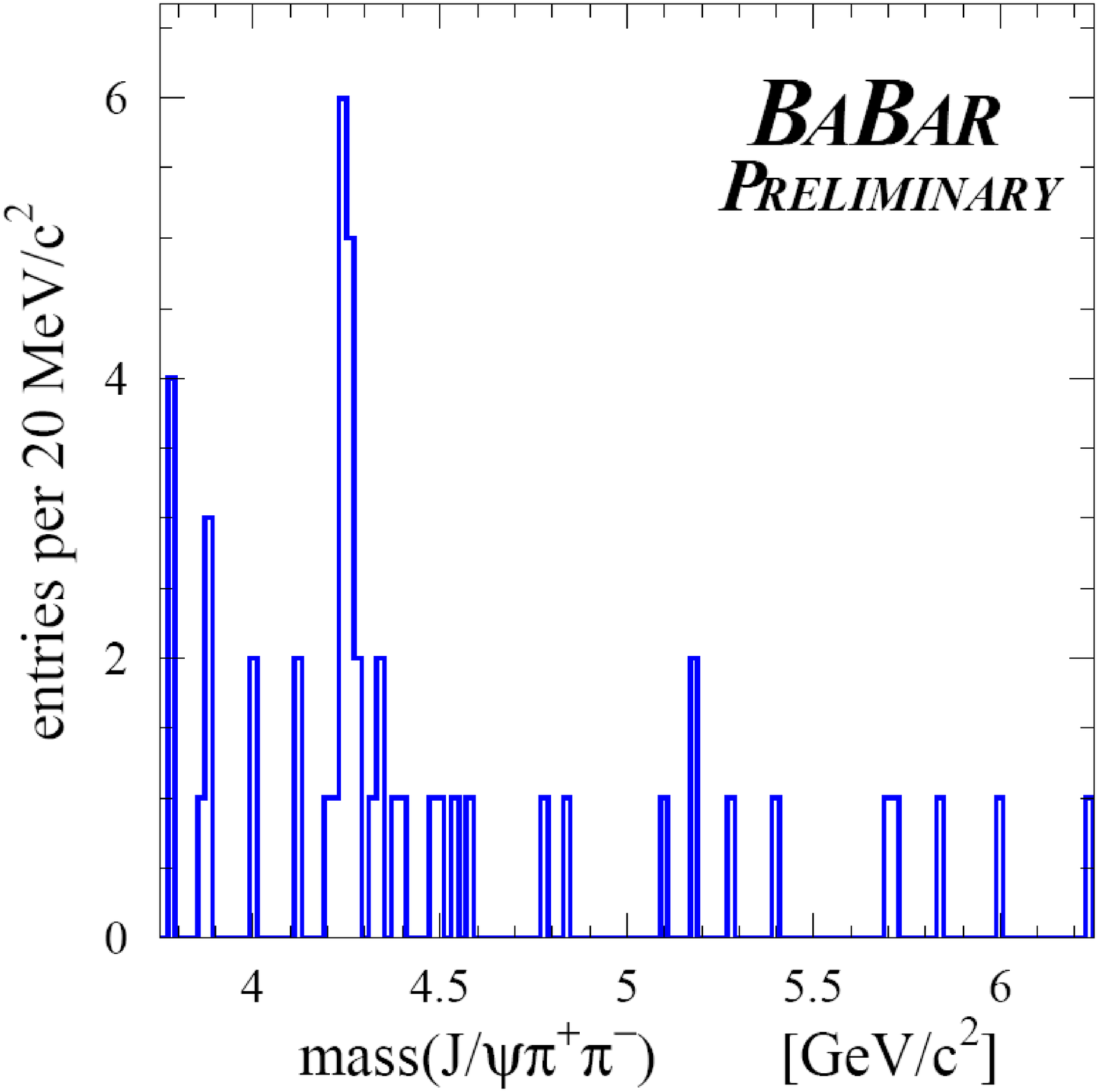}
 \end{center}
  \caption{The $\pi^+\pi^- J/\psi$ mass spectrum for the ISR sample, as measured
by BaBar in the analysis with the detection of the hard photon
radiated from an initial $e^+e^-$ collision.}
\label{YD2}
\end{figure}


The BaBar collaboration has studied the exclusive production of  the $D\bar{D}$
system ($D=D^0$ or $D^+$) through initial state radiation~\cite{Y4260_DD1}. As seen in Fig~\ref{YDD},
 the $D\bar{D}$ mass spectrum shows a clear $\psi(3770)$ signal and two further structures,
centered around 3.9 and 4.1 GeV/c$^2$. No evidence was found for $Y(4260)\to D\bar{D}$, leading to 
an upper limit 
$\frac{{\cal B}(Y(4260)\to D\bar{D})}{{\cal B}(Y(4260)\to \pi^+\pi^- J/\psi)} < 7.6$ (90~\% C.L.).
This number is over an order of magnitude smaller to compare with the value for the $\psi(3770)$
which makes the interpretation of $Y(4260)$ as a conventional $c\bar{c}$, rather doubtful.

The BaBar collaboration has also searched for the processes 
$e^+e^- \to (J/\psi\gamma\gamma)\gamma_{ISR}$
and $e^+e^- \to (J/\psi\pi^+\pi^-)\gamma_{ISR}$~\cite{Y4260_DD2},
where the hard photon radiated from an initial electron-positron collision
is  directly detected. In the latter final state the signal of  $Y(4260)$ was
observed~(Fig.~\ref{YD2}). 
Its mass and width are consistent with the the values originally reported by 
BaBar in~\cite{Y4260}.
In the $(J/\psi\gamma\gamma)\gamma_{ISR}$ final state, no events were found
in the $Y(4260)$ mass region in the $J/\psi\eta$, $J/\psi \pi^0$ and 
$\chi_{c2}\gamma$ distributions.

\begin{figure}[tb]
\begin{minipage}[b]{.5\linewidth}
\setlength{\unitlength}{1mm}
  \begin{picture}(70,50)
    \includegraphics[height=5.0cm,width=6cm]{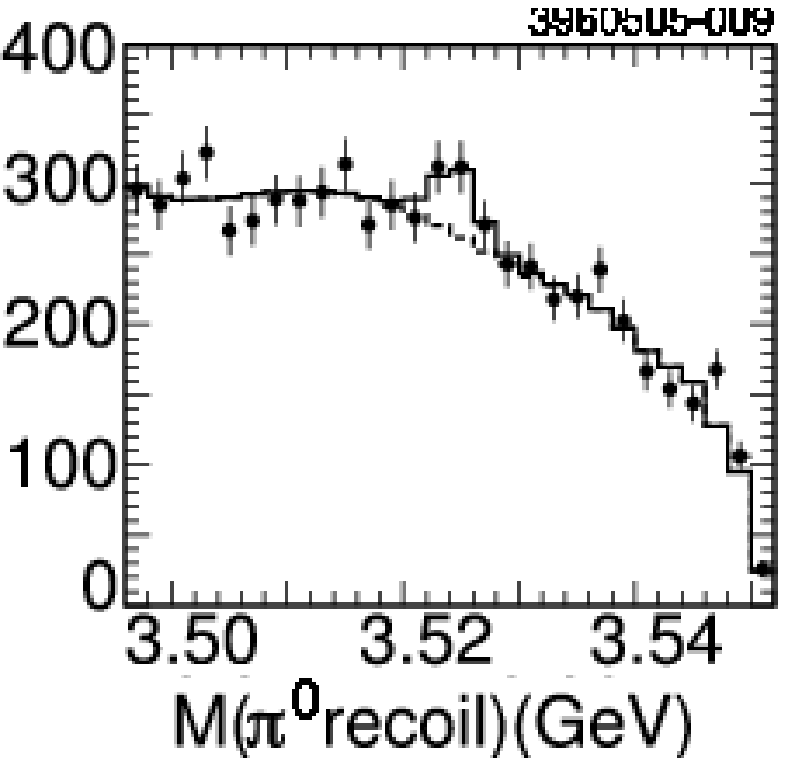}
  \put(-15,44){\bf (a)}
  \end{picture}
\end{minipage}\hfill
\begin{minipage}[b]{.5\linewidth}
\setlength{\unitlength}{1mm}
  \begin{picture}(70,50)
    \includegraphics[height=5.0cm,width=7.5cm]{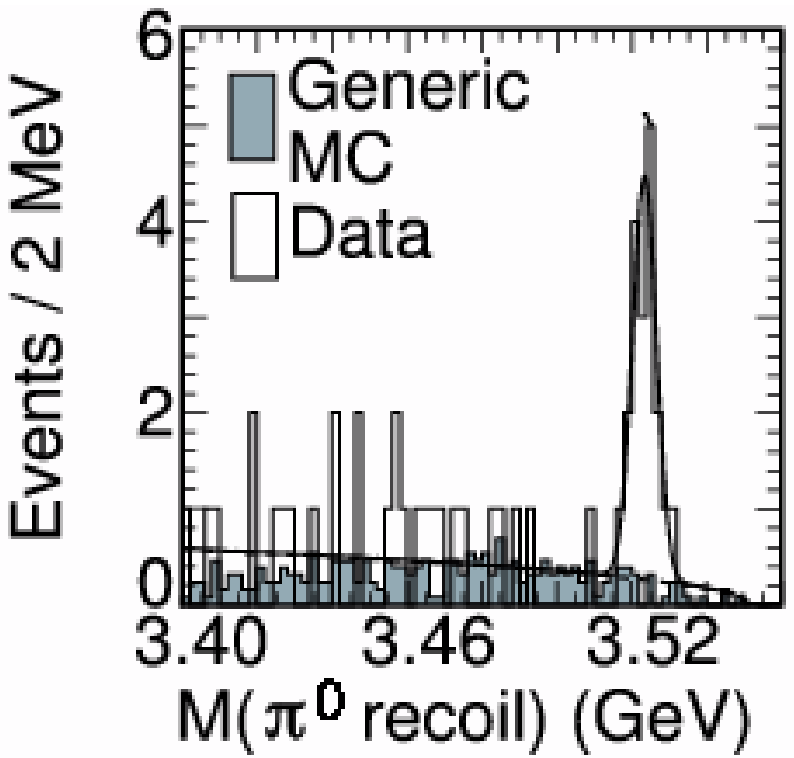}
  \put(-15,44){\bf (b)}
  \end{picture}
\end{minipage}
  \caption{The recoil mass against $\pi^0$ for {\bf (a)} inclusive (i.e. no reconstruction
of the $\eta_c$) and {\bf (b)} exclusive $\eta_c$ reconstruction 
in the reaction $\psi(2S)\to \pi^0 h_c \to (\gamma\gamma)(\gamma\eta_c)$, as measured by CLEO
collaboration.}
 \label{HCCLEO}
\end{figure}


\section{$h_c$}


The CLEO collaboration has observed the $h_c$ ($^1P1$) state of charmonium
in the reaction $\psi(2S)\to \pi^0 h_c \to (\gamma\gamma)(\gamma\eta_c)$~\cite{HCCLEO}.
The signal in the $\pi^0$ recoil mass was observed both for the inclusive reaction
(Fig.~\ref{HCCLEO} a)),
where the decay products 
of the $\eta_c$ are not identified, and for exclusive processes (Fig.~\ref{HCCLEO} b)),
 in which $\eta_c$
decays are reconstructed in seven hadronic decay channels ($\sim 10$~\% of all $\eta_c$ decays). 
The results of the inclusive and exclusive analyses were combined and yielded 
$M(h_c)=(3524.4\pm 0.6\pm 0.4)$~MeV/c$^2$ (in agreement with~\cite{FAUSTOV3}) and 
${\cal B}(\psi(2S)\to \pi^0 h_c)\times {\cal B}(h_c\to\gamma\eta_c) =(4.0\pm 0.8\pm 0.7)\times 10^{-4} $.
Together with the well known mass value of the $^3P_J$ centroid 
($<M(^3P_J)>=(3525.36\pm 0.06)$~MeV/c$^2$~\cite{PDG}), it has allowed to determine for the first
time the  hyperfine splitting for  the $P$ states of charmonium:
$\Delta M_{hf}(<M(^3P_J)> - M(^1P_1) = (+1.0 \pm 0.6 \pm 0.4)$ MeV/c$^2$.
This agrees well with the simplest calculations assuming the potential
composed of a vector Coulombic ($\sim r$) and a scalar confining ($\sim 1/r$)
terms. They are both spin independent and as a result the hyperfine splitting should be zero.
Larger values of the $\Delta M_{hf}$ could be accomodated only after the inclusion of 
the higher-order corrections~\cite{HCTHEOR1,HCTHEOR2}, which is not confirmed  by the CLEO measurement.

\begin{figure}[p]
\begin{center}
\centering\epsfig{figure=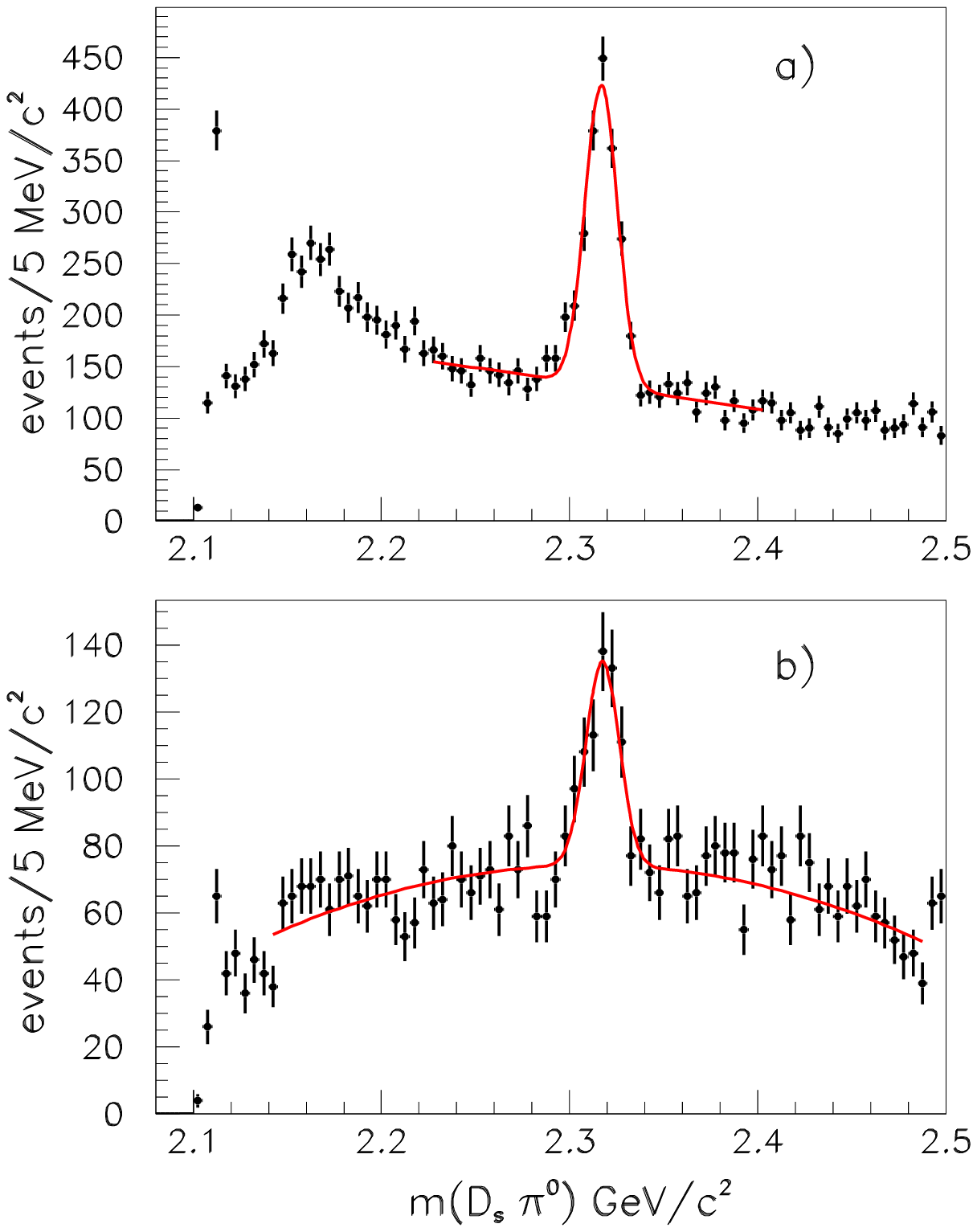,height=7.5cm}
\end{center}
\caption{The $D_s^+\pi^0$ mass distributions 
for {\bf (a)}
the decay $D_s^+\to K^+K^-\pi^+$ and {\bf (b)}
$D_s^+\to K^+K^-\pi^+\pi^0$, as measured by BaBar. The solid
curves represent the fits, described in~\cite{DSJ_BABAR1}.}
\label{FIG_DSJ_BABAR}
\begin{center}
\centering\epsfig{figure=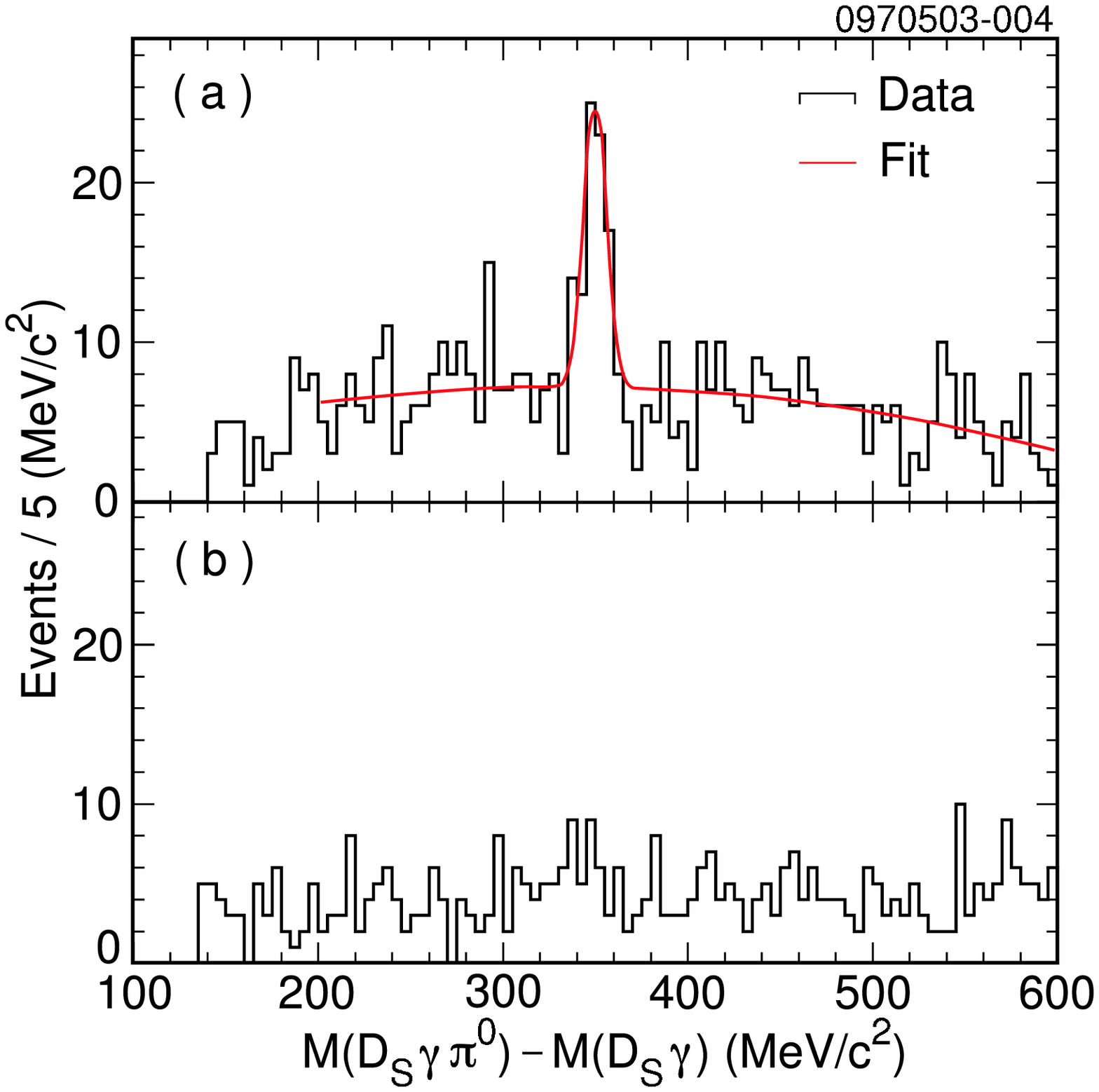,height=7.5cm}
\end{center}
\caption{The mass difference $\Delta M(D_s^{*}\pi^0) = M(D_s\gamma\pi^0) - M(D_s\gamma)$,
measured by the Cleo collaboration
for {\bf (a)} combinations where the $D_s\gamma$ system is consistent with $D_s^*$
decay and {\bf (b)} $D_s\gamma$ combinations selected from the $D_s^*$ mass sideband regions.}
\label{FIG_DSJ_CLEO}
\end{figure}


\section{$D_{sJ}$ mesons}


\begin{figure}[p]
\begin{center}
\centering\epsfig{figure=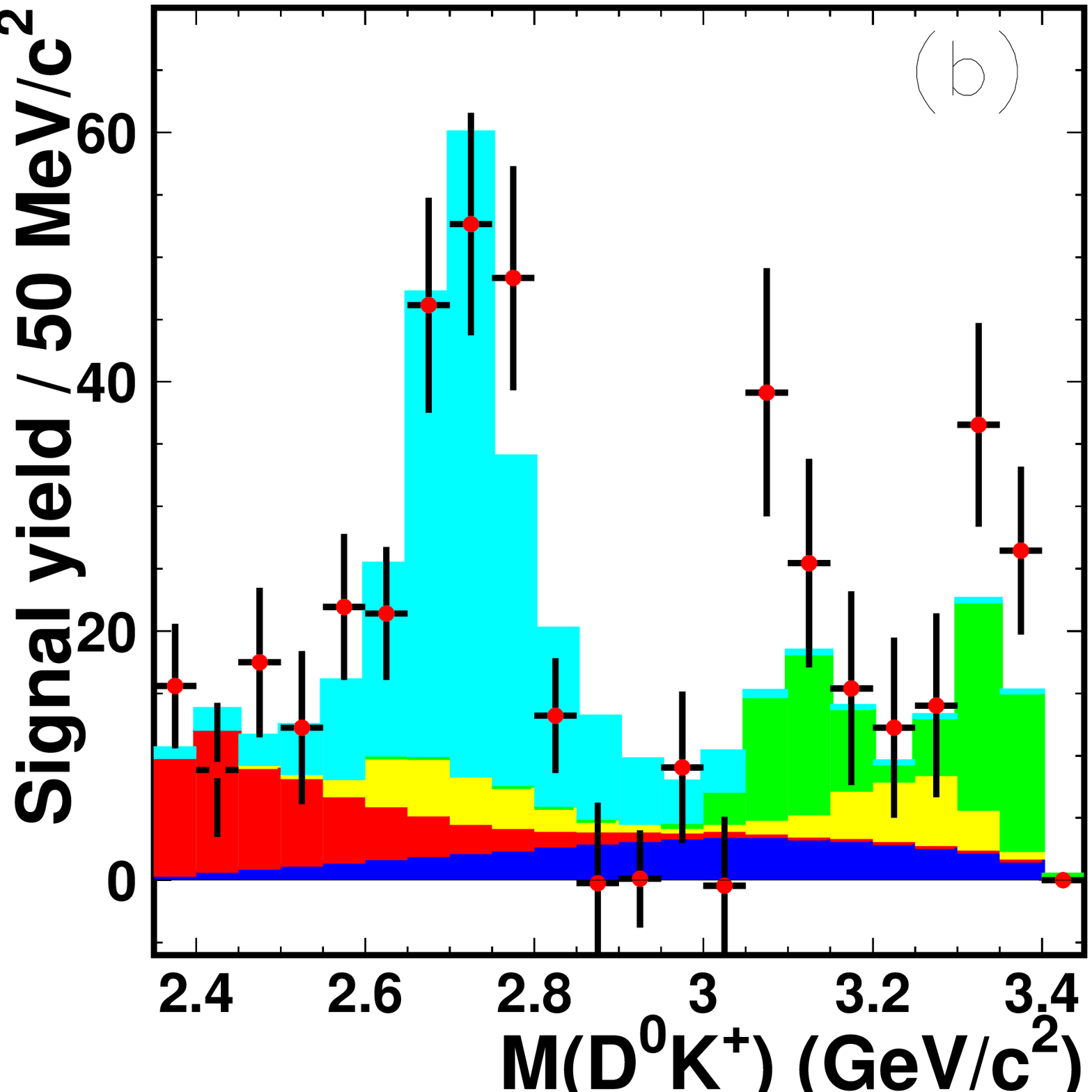,height=7cm}
\end{center}
\caption{$B^+\to\bar{D^0}D^0 K^+$ signal yield vs $M(D^0K^+)$ (data points) as 
measured by Belle. Additively superimposed histograms denote the contributions from $D_{sJ}(2700)$ 
(blue), $\psi(3770)$ (green), $\psi(4160)$ (yellow), threshold (red) and 
phase-space (navy blue)  components.}
\label{FIG_BRODZICK}
\begin{center}
\centering\epsfig{figure=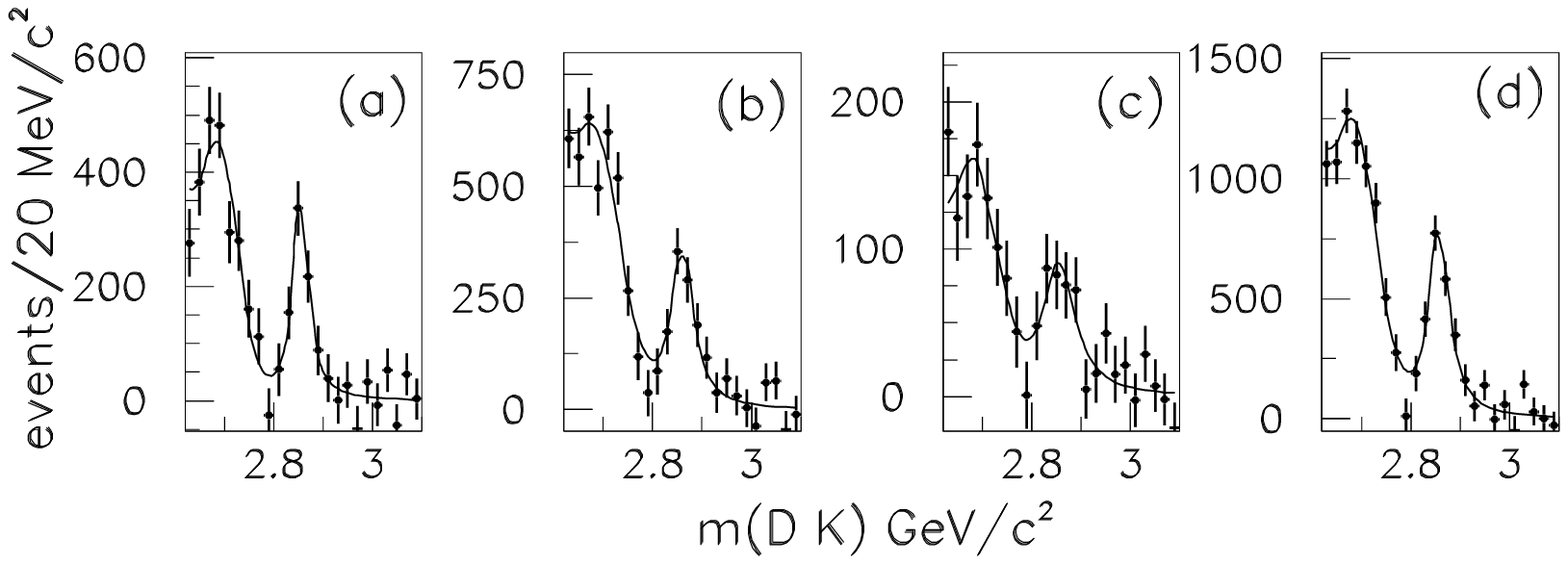,height=7cm,width=17cm}
\end{center}
\caption{Background subtracted $DK$ invariant mass distributions
measured by BaBar collaboration for 
{\bf (a)} $D^0(\to K^-\pi^+)K^+$, {\bf (b)} $D^0(\to K^-\pi^+\pi^0)K^+$,
{\bf (c)} $D^+(\to K^-\pi^+\pi^+)K^0_s$ and {\bf (d)} the sum of all modes 
in the 2.86 GeV/c$^2$ mass range. The curves are the fitted functions described 
in~\cite{DSJ_BABAR_286}.}
\label{FIG_DSJ_BABAR286}
\end{figure}

The symbol $D_{sJ}$ is often used to mark orbital excitations of the 
$c\bar{s}$ bound states. Four such $P$-wave mesons are  expected in 
the framework of potential models, inspired by the
heavy quark symmetry (HQS)~\cite{HQS1,HQS2}.
They can be naturally splitted in two doublets differing in the orbital
momentum of light degrees of freedom ($j_q$). The states with $j_q=3/2$ are 
predicted to be narrow and were identified as the
$D_{s1}(2536)$ (Argus)  and $D_{s2}(2573)$ (Cleo) in 1989 (1994),
respectively.
The mesons belonging to the $j_q=1/2$ are expected to be much wider i.e.
more difficult to observe.

Two candidates for such states were found in 2003.
First the BaBar collaboration provided the evidence for the state $D_{sJ}(2317)^+$
 (Fig.~\ref{FIG_DSJ_BABAR}),
decaying to $D_s^+\pi^0$~\cite{DSJ_BABAR1}. The observation by Cleo 
of the second state $D_{sJ}(2460)$ 
 (Fig.~\ref{FIG_DSJ_CLEO}),
in the decay to $D_s^{*+}\pi^0$~\cite{DSJ_CLEO1},
followed almost immediately. Both signals were found in the continuum processes 
$e^+e^-\to c\bar{c}$. There were soon confirmed by the Belle collaboration, together
with an additional evidence of their presence in $B$ meson exclusive decays~\cite{DSJ_BELLE1}.
Two other decay modes  to the final states $D_s\gamma$ and $D_s\pi^+\pi^-$ (implies a spin
of at least one) were observed for the $D_{sJ}(2460)^+$.

The discussed below, unexpected properties of both new mesons questioned the interpretation
of both new mesons as ($c\bar{s}$, $j_q=1/2$) bound states. 
First of all the widths of 
both the $D_{sJ}(2317)$
and $D_{sJ}(2460)$ turned out to be very small, consistent with the experimental 
resolution ($< 4.6$~MeV and $< 5.5$~MeV, respectively).
Also their masses, 
measured  to be below the $DK$ (D*K) thresholds, respectively,
appeared to be significantly lower to compare with HQS expectations, 
On the other side the study of angular distributions of the $D_{sJ}(2317)$ 
and $D_{sJ}(2460)$ 
decay products, performed by BaBar~\cite{DSJ_BABAR2} and Belle~\cite{DSJ_BELLE2},
 favoured strongly their spin-parity assignments $0^+$
and $1^+$, in agreement with HQS predictions.
This motivated a vigorous answer from the side of theorists, proposing several
exotic explanations for the two new mesons. In particular, the  
$D_{sJ}(2317)$  and $D_{sJ}(2460)$ were interpreted as
$D^{(*)}K$ molecules~\cite{DSJ_DK1,DSJ_DK2} or  the chiral doublers of 
the $D_s$ and $D_s^*$~\cite{DSJ_CHIRAL1,DSJ_CHIRAL2}.
Assuming that  current mass predictions of HQS are wrong (they can in fact be shifted
to lower values, in the presence of a strong $S$ wave coupling to $D K^{(*)}$), both new $D_{sJ}$
mesons could be comfortably interpreted  as coventional $(c\bar{s})$ states.
Provided that their predicted masses may be lowered below the respective $D^{(*)}K$
thresholds, the narrow widths of the $D_{sJ}(2317)$  and $D_{sJ}(2460)$ are naturally explained.
These low masses would clearly allow the observed electromagnetic and isospin-violating
decays of the the two states to be pronounced. Thus, the two new $D_{sJ}$
mesons can be interpreted as conventional states 
$D_{s0}^*$ and
$D_{s1}$~(\cite{FAZIO1}--\cite{FAZIO3}).

Yet another charm-strange meson, marked as $D_{sJ}(2700)$ and produced in $B^+\to \bar{D^0}D_{sJ}$, $D_{sJ}\to D^0 K^+$
was observed by Belle~\cite{DSJ_BRODZICK} (Fig.~\ref{FIG_BRODZICK}).
This state has a mass of $M=(2715\pm 11 ^{+11}_{-14})$ MeV/c$^2$ and a width
$\Gamma=(115\pm 20^{+36}_{-32})$ MeV and its signal corresponds to $182\pm 30$ events.
The study of $D_{sJ}(2700)$ helicity angle distributions strongly favours
the spin parity assignment of $1^-$.

Recent observations concerning the $D_{sJ}$ family are completed by the
study of three inclusive processes
$e^+e^-\to D^0 K^+ X, D^0\to K^-\pi^+$,
$e^+e^-\to D^0 K^+ X, D^0\to K^-\pi^+\pi^0$ and
$e^+e^-\to D^+ K^0_s X, D^+\to K^-\pi^+\pi^+$ performed by BaBar~\cite{DSJ_BABAR_286}.
The distributions of $DK$ invariant mass (Fig.~\ref{FIG_DSJ_BABAR286}) show a clear signal
of a new charm-strange meson, marked as $D_{sJ}(2860)$, with a mass of
$M=(2856.6\pm 1.5\pm 5.0)$ MeV/c$^2$ and a width $\Gamma =(47\pm 7\pm 10)$ MeV.
The decay to two pseudoscalars implies a natural spin-parity for this state
($0^+, 1^-,...$) and the value $J^P=3^-$ is predicted in~\cite{FAZIO4}. 
  According to~\cite{DSJ286_INT}, the $D_{sJ}(2860)$ could be 
a radial  excitation of $D_{sJ}^*(2317)$. However, other assignments cannot be ruled out.
Moreover, a second broad enhancement is observed around 2.69 GeV/c$^2$ (Fig.~\ref{FIG_DSJ_BABAR286}).
This state was temporarily marked as $X(2690)^+$ and clearly further inputs are 
necessary in order to understand its origin.
Its mass was determined to be 
$M=(2688\pm 4\pm 3)$~MeV/c$^2$ and a width $\Gamma = (112\pm 7\pm 36)$~MeV.
It would be very interesting to check if there is any association between
the $D_{sJ}(2700)$ and $X(2690)$.

\begin{figure}[p]
\setlength{\unitlength}{1mm}
\begin{center}
\begin{picture}(230,60)(-12,0)
\put(-1,20){\rotatebox{90}{\tiny\bf events/(10~{\rm MeV}/$c^2$)}}
\put(43,1){{\large $M(\Lambda_c^+\pi)-M(\Lambda_c^+)$ [GeV/$c^2$]}}
\put(25,45){\large $\Lambda_c^+\pi^-$}
\put(63,45){\large $\Lambda_c^+\pi^0$}
\put(103,45){\large $\Lambda_c^+\pi^+$}
\centering\epsfig{figure=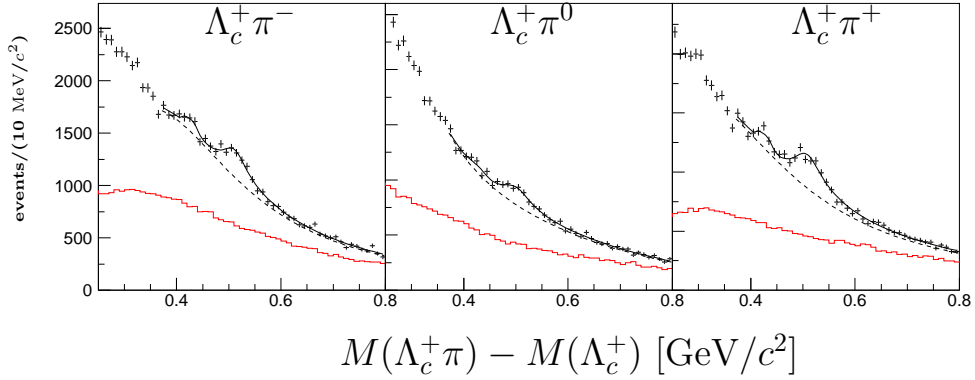,height=6cm}
\end{picture}
\end{center}
\caption{$M(\Lambda_c^+\pi)-M(\Lambda_c)$ distributions of the selected
$\Lambda_c^+\pi^-$ (left), $\Lambda_c^+\pi^0$ (middle), and $\Lambda_c^+\pi^+$ (right)
 combinations. Data from the $\Lambda_c^+$ signal window (points with error bars) and 
normalized sidebands (histograms) are shown, together with the fits (solid curves) and
 their combinatorial background components (dashed).}
\label{FIG_MIZUK}
\end{figure}

\begin{table}[p]
\caption{Parameters of the baryons $\Sigma_c(2800)^0$, $\Sigma_c(2800)^+$ and $\Sigma_c(2800)^{++}$ as
measured by Belle.}
\begin{center}
\begin{tabular}{l|cccc}
State & $\Delta M$ [{\rm MeV}/c$^2$] & Width [{\rm MeV}] & Yield/$10^3$ & Significance ($\sigma$)   \\ 
\hline
$\Sigma_c(2800)^0$    &  $515.4^{+3.2+2.1}_{-3.1-6.0} $ & $61^{+18+22}_{-13-13}$ 
 & $2.24^{+0.79+1.03}_{-0.55-0.50}$ & ~8.6 \\
$\Sigma_c(2800)^+$    &  $~505.4^{+5.8+12.4}_{-4.6-2.0}$ & $62^{+37+52}_{-23-38}$ 
 & $1.54^{+1.05+1.40}_{-0.57-0.88}$ & ~6.2 \\
$\Sigma_c(2800)^{++}$ &  $514.5^{+3.4+2.8}_{-3.1-4.9} $ & $75^{+18+22}_{-13-11}$ 
 & $2.81^{+0.82+0.71}_{-0.60-0.49}$ & 10.0 \\
\hline
\end{tabular}
\end{center}
\label{TABLE_MIZUK}
\caption{Parameters of the baryons $\Lambda_c(2880)^+$ and  $\Lambda_c(2940)^+$, as determined by BaBar and Belle.}
\begin{center}
\begin{tabular}{l|cccc}
State & Expt. & Mass [{\rm MeV}/c$^2$] & Width [{\rm MeV}] & Yield (events)   \\ 
\hline
$\Lambda_c(2880)^+$ & BaBar  &  $2881.9 \pm 0.1 \pm 0.5 $      & $5.8\pm 1.5\pm 1.1$ 
 & $2800\pm 190$  \\
$\Lambda_c(2880)^+$ & Belle  &  $2881.2 \pm 0.2^{+0.4}_{-0.3}$ & $5.5^{+0.7}_{-0.3}\pm 0.4$ 
 & $880\pm 50\pm 40$  \\
\hline
$\Lambda_c(2940)^+$ & BaBar  &  $2939.8 \pm 1.3 \pm 1.0 $      & $17.5\pm 5.2\pm 5.9$ 
 & $2280\pm 310$  \\
$\Lambda_c(2940)^+$ & Belle  &  $2937.9 \pm 1.0^{+1.8}_{-0.4}$ & $10\pm 4\pm 5$ 
 & $210^{+70+100}_{-40-60}$  \\
\hline
\end{tabular}
\end{center}
\label{TABLE_LC2940}
\end{table}


\section{$\Sigma_{c}(2800)$}


The Belle collaboration  has provided 
the first evidence~\cite{MIZUK}
 for an isotriplet of excited charmed baryons $\Sigma_c(2800)$
decaying into the $\Lambda_c^+\pi^-$, $\Lambda_c^+\pi^0$ and $\Lambda_c^+\pi^+$ final 
states.
As shown in Fig.~\ref{FIG_MIZUK}, clear enhancements around 0.51 GeV/c$^2$ are seen in the
distributions of the mass difference 
$\Delta M (\Lambda_c^+\pi)= M(\Lambda_c^+\pi) - M(\Lambda_c^+)$
for the $\Lambda_c^+\pi^-$, $\Lambda_c^+\pi^0$, and $\Lambda_c^+\pi^+$ combinations.
The mass differences  $\Delta M$ together with the widths
of the states $\Sigma_c(2800)$ are collected in Table~\ref{TABLE_MIZUK}.
These states are identified as the members of the predicted
$\Sigma_{c2}$, $J^P=3/2^-$ isospin triplet~\cite{MIZUK_THEOR}.
 The enhancement near $\Delta M=0.43$ GeV/c$^2$
(cf Fig.~\ref{FIG_MIZUK}),
in the spectra corresponding to 
$\Lambda_c^+\pi^-$ and 
$\Lambda_c^+\pi^+$ combinations, is attributed to feed-down from the decay 
$\Lambda_c(2880)^+\to \Lambda_c^+\pi^+\pi^-$, as verified by reconstructing
$\Lambda_c(2880)$ in the data. 

\begin{figure}[ph]
 \begin{center}
  \includegraphics[height=7.0cm]{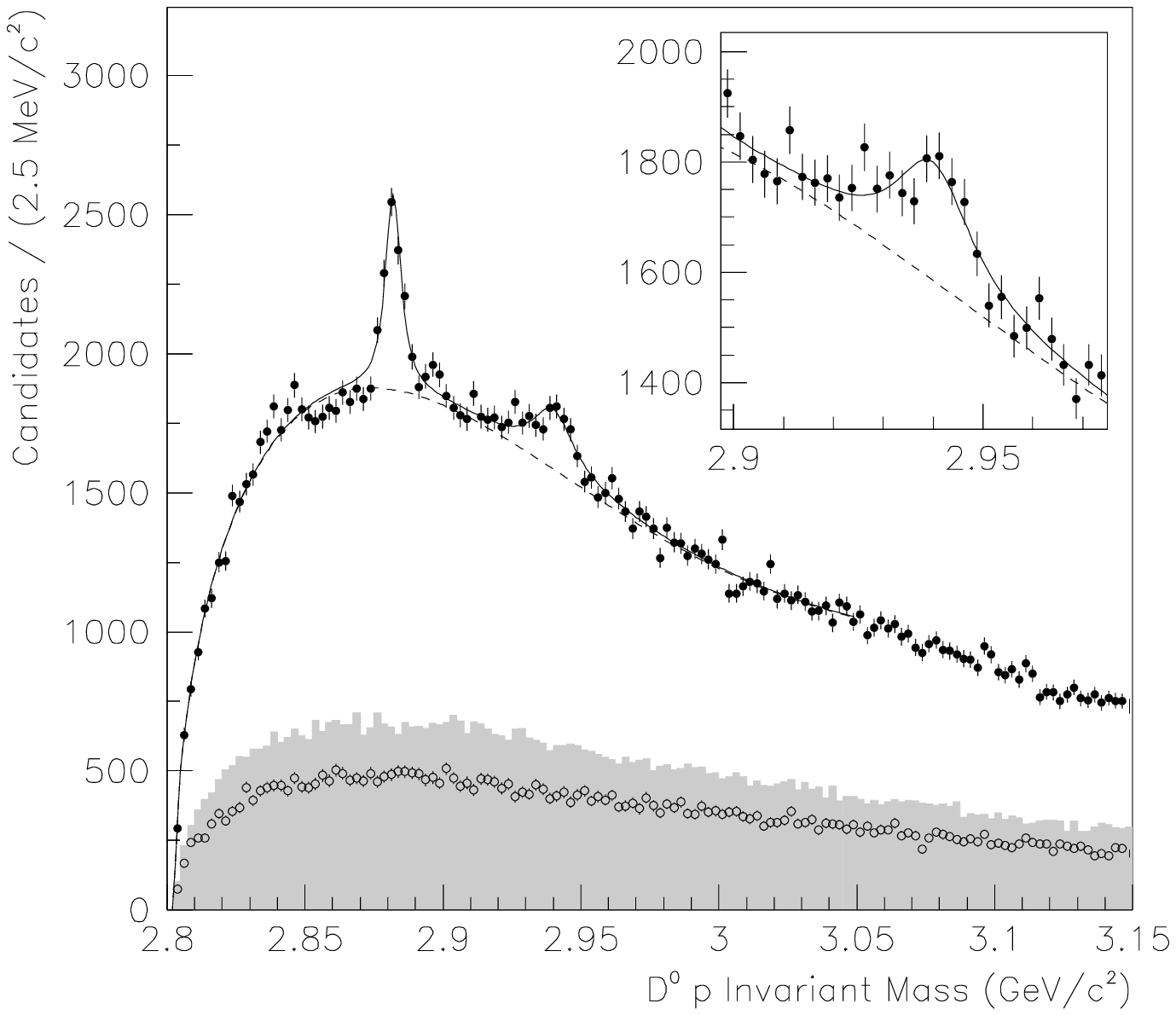}
 \end{center}
  \caption{Invariant mass  distribution of $p D^0$  pairs, as measured by BaBar collaboration
(data points). The shaded histogram and open circles correspond to the $D^0$ mass sidebands and
wrong-sign $p\overline{D^0}$ candidates, respectively. The inset shows the $pD^0$
mass spectrum in the range 2.9--2.975 GeV/c$^2$.}
\label{FIGBAB_LC2940}
\setlength{\unitlength}{1mm}
\begin{center}
\begin{picture}(100,75)
\put(4,40){\rotatebox{90}{ $N / 2.5$ MeV/c$^2$}}
\put(35,+5){{$m(\Lambda_c^+\pi^+\pi^-)$ [GeV/$c^2$]}}
\centering\epsfig{figure=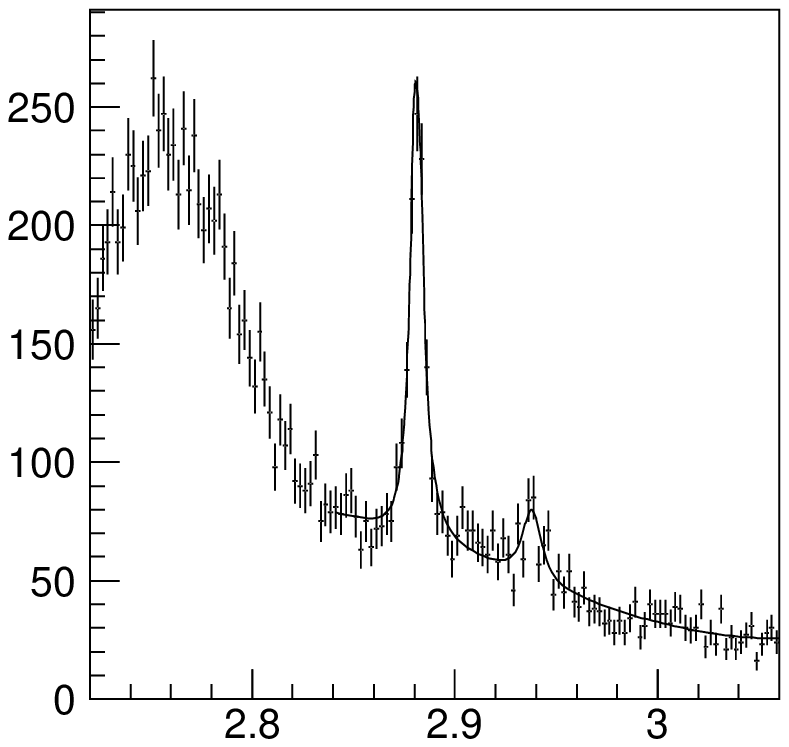,height=8.5cm,width=10cm}
\end{picture}
\end{center}
\caption{The invariant mass distribution of the $\Lambda_c^+\pi^+\pi^-$
combinations as measured by the Belle collaboration. 
The plot corresponds to the $\Sigma_c(2455)$ mass peak of the $\Lambda_c\pi^{\pm}$ 
combinations. The signals of $\Lambda_c(2765)^+$, $\Lambda_c(2880)^+$
and $\Lambda_c(2940)^+$ can be clearly distinguished.}
\label{FIGBEL_LC2940}
\end{figure}



\section{$\Lambda_c(2940)^+$ and $\Lambda_c(2880)^+$}


The charmed  baryon $\Lambda_c(2940)^+$ was first observed by the 
BaBar collaboration in the $p D^0$ final state~\cite{LC2940_BABAR} (Fig.~\ref{FIGBAB_LC2940}).
The signal at 2.88 GeV/c$^2$ is due to the decay $\Lambda_c(2880)^+\to p D^0$.
This comprises the first observation of the above decay channel 
 (the baryon $\Lambda_c(2880)^+$ was first seen by CLEO in the final state
$\Lambda_c^+\pi^+\pi^-$~\cite{LC2880_CLEO}). The search for a doubly-charged partner of the
$\Lambda_c(2940)^+$, performed by BaBar in the final state $p D^+$, gave negative results~\cite{LC2940_BABAR}.

The Belle collaboration has recently reported the evidence for another decay mode
$\Lambda_c(2940)^+\to \Sigma_c(2455)^{0,++}\pi^{\pm}$~\cite{LC2940_BELLE}
(Fig.~\ref{FIGBEL_LC2940}). The study of angular distributions of the decay
$\Lambda_c(2880)^+\to \Sigma_c^{0,++}\pi^{\pm}$ strongly favours a $\Lambda_c(2880)^+$
spin assignment of $\frac{5}{2}$ over $\frac{3}{2}$ and $\frac{1}{2}$~\cite{LC2940_BELLE}.

The parameteters of both $\Lambda_c(2880)^+$
and $\Lambda_c(2940)^+$ measured by Belle and BaBar, are in good overall agreement
(Table~\ref{TABLE_LC2940}).
 


\section{$\Xi_{cx}(2980)$ and $\Xi_{cx}(3077)$}


\begin{figure}[th]
\setlength{\unitlength}{1mm}
\begin{picture}(190,50)
\put(13,44){{\bf (a) }}
\begin{minipage}[b]{.5\linewidth}
\begin{center}
\epsfig{figure=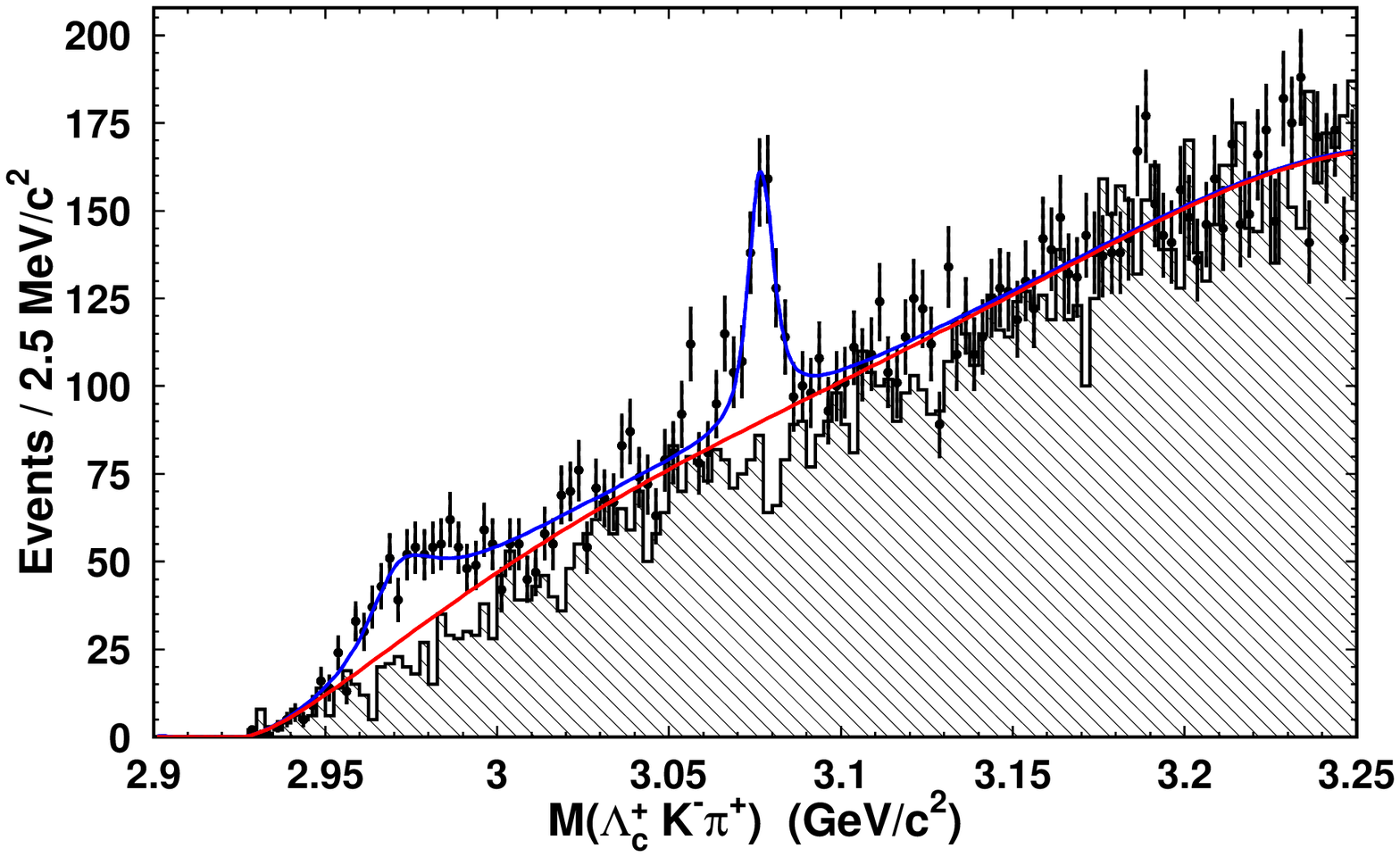,width=\linewidth,height=5.5cm}
\end{center}
\end{minipage}\hfill
\begin{minipage}[b]{.5\linewidth}
\begin{center}
\epsfig{figure=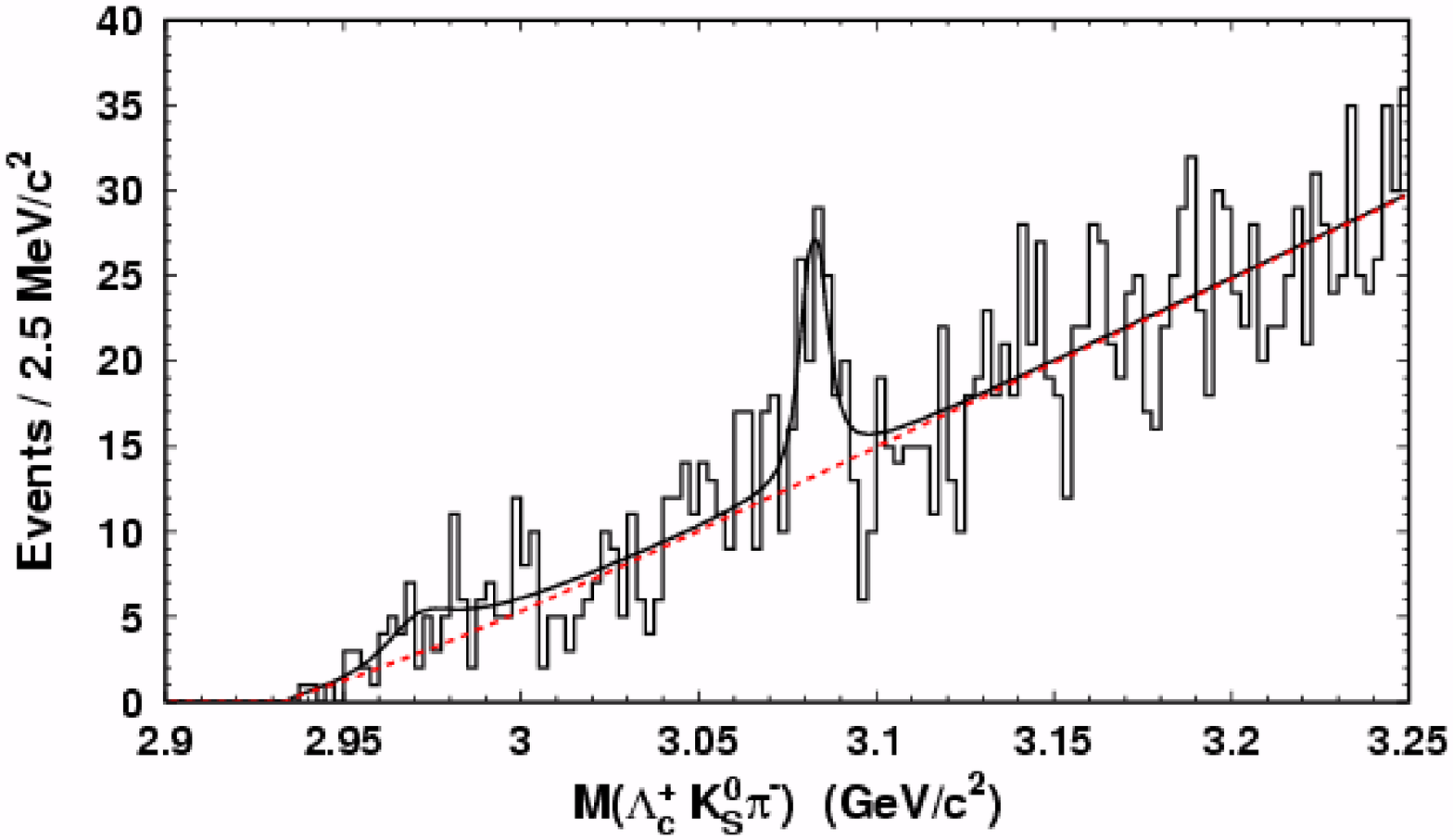,width=\linewidth,height=5.3cm}
\put(86,44){{\bf (b) }}
\end{center}
\end{minipage}
\end{picture}
\caption{ {\bf (a)}: $M(\Lambda_c^+ K^-\pi^+)$ distribution (points with 
error bars) together with the fit (solid curve). The dashed region
represents the background component corresponding to the wrong-sign
combinations $\Lambda_c^+K^+\pi^-$. 
{\bf (b)}: $M(\Lambda_c^+ K^0_s\pi^+)$ distribution (points) together 
with the overlaid fitting curve. Both spectra were measured by the Belle
collaboration.
\label{XICX1}}
\end{figure}

\begin{figure}[hp]
 \begin{center}
  \includegraphics[height=16.0cm]{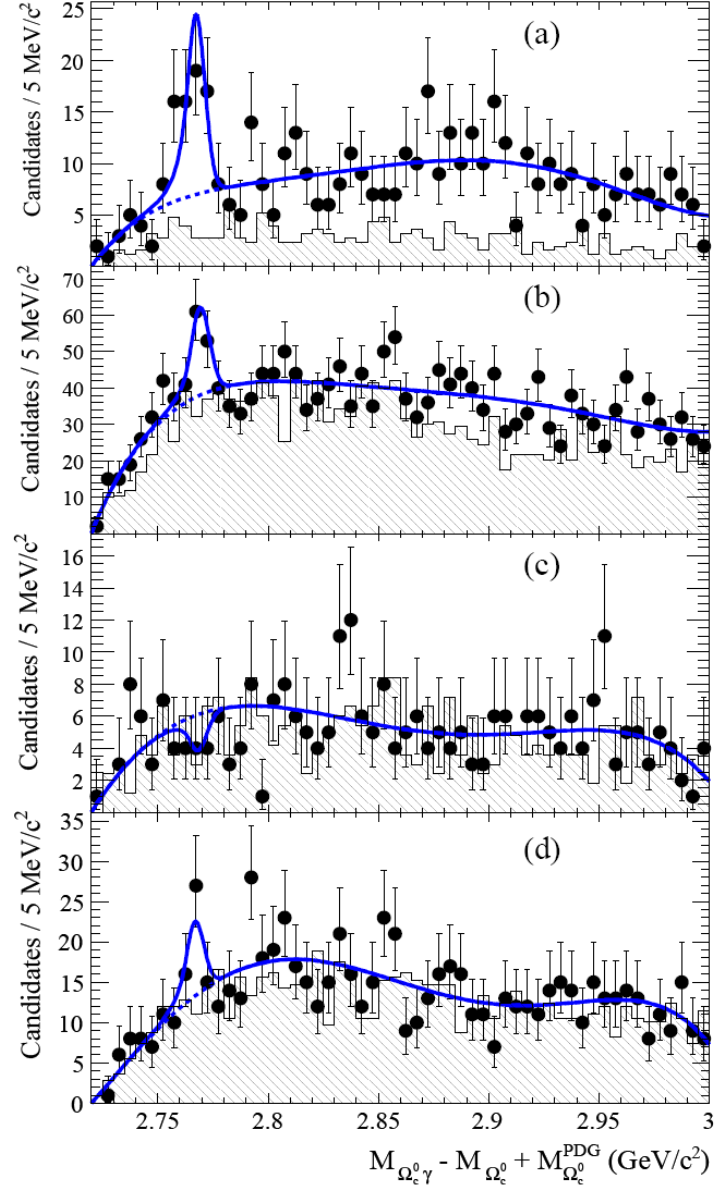}
 \end{center}
  \caption{The invariant mass  distributions of $\Omega_c^{*0}\to\Omega_c^0\gamma$ candidates
with $\Omega_c^0$ reconstructed by BaBar in the decay modes 
{\bf (a)} $\Omega^-\pi^+$, {\bf (b)} $\Omega^-\pi^+\pi^0$, {\bf (c)} $\Omega^-\pi^+\pi^-\pi^+$
and {\bf (d)} $\Xi^- K^-\pi^+\pi^+$.
The points with error bars represent the data, the dashed line corresponds to the combinatorial 
background and the solid line is the sum of signal and background. The shaded histograms correspond 
to the mass distribution expected from the mass sideband of $\Omega_c^0$.}
\label{OMEGAC}
\end{figure}

In the beginning of this year, the Belle collaboration 
reported the first observation of two baryons~\cite{XICX_BELLE}, denoted as 
$\Xi_{cx}(2980)^{+}$ and
$\Xi_{cx}(3077)^{+}$ and 
decaying into $\Lambda_c^+ K^-\pi^+$ (Fig.~\ref{XICX1}(a)).
The existence of both new particles were quickly confirmed by BaBar~\cite{XICX_BABAR}.
 Assuming that these
states carry charm and strangeness, the above observation 
would comprise the first example of a baryonic decay in which the initial $c$ and $s$ 
quarks are carried away by two different final state particles.
Most naturally, these two states would be interpreted as excited charm-strange
baryons $\Xi_c$. 
This interpretation is strengthened by the positive results of the search
for neutral isospin related partners of the above states (Fig.~\ref{XICX1}(b)), performed by Belle in 
$\Lambda_c^+ K^0_s\pi^-$ final state~\cite{XICX_BELLE}. It yielded an evidence of the 
$\Xi_{cx}(3077)^0$ 
 together with a broad enhancement 
near the threshold i.e. in the mass range corresponding to the $\Xi_{cx}(2980)^0$.
The  preliminary parameters of the states 
$\Xi_{cx}(2980)^+$ and
$\Xi_{cx}(3077)^+$ are collected in Table~\ref{TABLE_CHISTOV}.

In the $\Lambda_c^+ K^-\pi^+ (\pi^+)$ final state, the SELEX collaboration~\cite{SELEX} reported
the observation of two double charmed baryon: $\Xi_{cc}^+$ with a mass of 3520 MeV/c$^2$ and 
$\Xi_{cc}^{++}$ with a mass of 3460 MeV/c$^2$. The studies by Belle~\cite{XICX_BELLE}
and BaBar~\cite{XICX_BABAR2}  
show no evidence for these states.
The BaBar collaboration estimated the following 95~\% C.L. upper limits
on the ratio of production cross-sections:
$\sigma(\Xi_{cc}^+)\times {\cal B}(\Xi_{cc}^+\to \Lambda_c^+ K^-\pi^+)/\sigma(\Lambda_c^+) < 2.7 \times 10^{-4}$ and 
$\sigma(\Xi_{cc}^{++})\times {\cal B}(\Xi_{cc}^{++}\to \Lambda_c^+ K^-\pi^+\pi^+)/\sigma(\Lambda_c^+) < 4.0 \times 10^{-4}$ 
(estimated for $p^*(\Lambda_c) > 2.3$ GeV/c,   where $p^*$ denotes the CMS momentum of the $\Lambda_c$). 
The Belle collaboration studied only the single-charged state which yielded
$\sigma(\Xi_{cc}^+)\times {\cal B}(\Xi_{cc}^+\to \Lambda_c^+ K^-\pi^+)/\sigma(\Lambda_c^+) < 1.5 \times 10^{-4}$
(90~\% C.L.; $p^*(\Lambda_c) > 2.5$ GeV/c).

\begin{table}[tbh]
\caption{Parameters of the  new charm-strange baryons $\Xi_{cx}(2980)^{+,0}$ and $\Xi_{cx}(3077)^{+,0}$}
\vspace{0.4cm}
\begin{tabular}{l|ccccc}
State & Expt. & Mass  & Width  & Yield  & Signif. \\ 
 & &({\rm MeV/c}$^2$)  &({\rm MeV})& (events) & ($\sigma$) \\ 
\hline
$\Xi_{cx}(2980)^+$  & BaBar & $2967.1\pm 1.9\pm 1.0$ & $23.6\pm 2.8\pm 1.3$ & $284\pm 45\pm 46$ & 7.0 \\
$\Xi_{cx}(2980)^+$  & Belle & $2978.5\pm 2.1\pm 2.0$ & $43.5\pm 7.5\pm 7.0$ & $405.3\pm 50.7$ & 5.7 \\
$\Xi_{cx}(3077)^+$  & BaBar & $3076.4\pm 0.7\pm 0.3$ & $~6.2\pm 1.6\pm 0.5$ & $204\pm 35\pm 12$ & 8.6 \\
$\Xi_{cx}(3077)^+$  & Belle & $3076.7\pm 0.9\pm 0.5$ & $~6.2\pm 1.2\pm 0.8$ & $326.0\pm 39.6$ & 9.2 \\
\hline
$\Xi_{cx}(2980)^0$  & Belle & $2977.1\pm 8.8\pm 3.5$ & $~43.5$ (fixed)      & $42.3\pm 23.8$ & 1.5 \\
$\Xi_{cx}(3077)^0$  & Belle & $3082.8\pm 1.8\pm 1.5$ & $~5.2\pm 3.1\pm 1.8$ & $67.1\pm 19.9$ & 4.4 \\
\hline
\end{tabular}
\label{TABLE_CHISTOV}
\end{table}



\section{$\Omega_{c}^{*0}$}


The baryon $\Omega_{c}^{*0}$ was observed by the BaBar collaboration in the 
radiative decay $\Omega_c^0\gamma$~\cite{OMEGAC_BABAR}.
This state was the last  singly-charm baryon having zero orbital momentum, 
remaining to be experimentally detected. 
The $\Omega_c^0$ was reconstructed in the decays to the final states
$\Omega^-\pi^+$, $\Omega^-\pi^+\pi^0$, $\Omega^-\pi^+\pi^-\pi^+$ and
$\Xi^- K^-\pi^+\pi^+$ (Fig.~\ref{OMEGAC}).
The mass difference between $\Omega_c^{*0}$ and $\Omega_c^{0}$
was measured to be $\Delta M = 70.8\pm 1.0 \pm 1.1$~MeV/c$^2$.
This agrees with the theoretical prediction in~\cite{OMEGAC_TH1,FAUSTOV2}
and is below that described in~\cite{OMEGAC_TH2}.


\section{Summary}


The charm physics has many features of the Sleeping Beauty.
After the initial publicity of the times of November revolution,
it remained a calm field  aimed at filling the columns of Particle
Data Group booklets with new or more accurate 
cross-sections, branching ratios, lifetimes etc.
It seems that $B$ factories acted like a prince who  kissed the
Sleeping Beauty and waked her up right in the beginning of this
century.
The discovery of plethora of new charmed states has revitalized
the charm physics and triggered many new theoretical ideas.
Since the $B$ factories are still collecting enormous samples of data,
it is rather likely that some new exciting and charming discoveries are
just around the corner.


\bigskip
I am very grateful to the organizers of the HQL2006 Conference for their
support and all efforts in making this venue successful.
Special thanks to Prof. S.Paul.

\vskip 1.5cm

\end{document}

%% file: econfmacros.tex
\def\beq{\begin{equation}}
\def\eeq#1{\label{#1}\end{equation}}
\def\eeqn{\end{equation}}

\def\beqa{\begin{eqnarray}}
\def\eeqa#1{\label{#1}\end{eqnarray}}
\def\eeqan{\end{eqnarray}}

\let\bar=\overbar

\def\Dslash{\not{\hbox{\kern-4pt $D$}}}
\def\dslash{\not{\hbox{\kern-2pt $\del$}}}

\def\msb{{\bar{\ssstyle M \kern -1pt S}}}

